\documentclass[aps,showpacs,prx,superscriptaddress,preprintnumbers,amsmath,amssymb,twocolumn]{revtex4-2}
\usepackage[utf8]{inputenc}
\usepackage[T1]{fontenc}
\usepackage{graphicx}
\usepackage{dcolumn}
\usepackage{appendix}
\usepackage{amsmath}
\usepackage{bm}
\usepackage[dvipsnames]{xcolor}
\usepackage{multirow}

\newcommand{\angstrom}{\text{\normalfont\AA}}

\newcommand{\tPsi}{\big\lvert\Tilde{\Psi}(t)\big\rangle}
\newcommand{\toPsi}{\big\lvert\Tilde{\Psi}(t_0)\big\rangle}

\newcommand{\chiV}{\lvert \bar{\chi}_v^{(q)}(t,t_0) \rangle} 
\newcommand{\chiVfund}{\lvert \bar{\chi}_v^{(q=1)}(t,t_0) \rangle}

\usepackage{subfigure}
\usepackage{bbm}
\usepackage{enumerate}

\usepackage[
bookmarks=true,
colorlinks,
linkcolor=black,
urlcolor=black,
citecolor=black,
plainpages=false,
pdfpagelabels,
final,
breaklinks=true
]{hyperref}

\usepackage{titlesec}

\usepackage{array}

\usepackage{physics}

\begin{document}
	\allowdisplaybreaks
	
	\title{Non-classical states of light after high-harmonic generation in semiconductors: a Bloch-based perspective}
	
	\author{J.~Rivera-Dean}
	\email{javier.rivera@icfo.eu}
	\affiliation{ICFO -- Institut de Ciencies Fotoniques, The Barcelona Institute of Science and Technology, 08860 Castelldefels (Barcelona)}
	
	\author{P. Stammer}
	\affiliation{ICFO -- Institut de Ciencies Fotoniques, The Barcelona Institute of Science and Technology, 08860 Castelldefels (Barcelona)}
	
	\author{A. S. Maxwell}
	\affiliation{Department of Physics and Astronomy, Aarhus University, DK-8000 Aarhus C, Denmark}
	
	\author{Th. Lamprou}
	\affiliation{Foundation for Research and Technology-Hellas, Institute of Electronic Structure \& Laser, GR-70013 Heraklion (Crete), Greece}
	\affiliation{Department of Physics, University of Crete, P.O. Box 2208, GR-70013 Heraklion (Crete), Greece}
	
	\author{A. F. Ordóñez}
	\affiliation{ICFO -- Institut de Ciencies Fotoniques, The Barcelona Institute of Science and Technology, 08860 Castelldefels (Barcelona)}
	
	\author{E. Pisanty}
	\affiliation{Department of Physics, King's College London, WC2R 2LS London, United Kingdom}
	
	\author{P. Tzallas}
	\affiliation{Foundation for Research and Technology-Hellas, Institute of Electronic Structure \& Laser, GR-70013 Heraklion (Crete), Greece}
	\affiliation{ELI-ALPS, ELI-Hu Non-Profit Ltd., Dugonics tér 13, H-6720 Szeged, Hungary}
	
	\author{M. Lewenstein}
	\email{maciej.lewenstein@icfo.eu}
	\affiliation{ICFO -- Institut de Ciencies Fotoniques, The Barcelona Institute of Science and Technology, 08860 Castelldefels (Barcelona)}
	\affiliation{ICREA, Pg. Llu\'{\i}s Companys 23, 08010 Barcelona, Spain}
	
	\author{M. F. Ciappina}
	\email{marcelo.ciappina@gtiit.edu.cn}
	\affiliation{Physics Program, Guangdong Technion--Israel Institute of Technology, Shantou, Guangdong 515063, China}
	\affiliation{Technion -- Israel Institute of Technology, Haifa, 32000, Israel}
	\affiliation{Guangdong Provincial Key Laboratory of Materials and Technologies for Energy Conversion, Guangdong Technion – Israel Institute of Technology, Shantou, Guangdong 515063, China}

	\date{\today}
	\begin{abstract}
		High-harmonic generation has emerged as a pivotal process in strong-field physics, yielding extreme ultraviolet radiation and attosecond pulses with a wide range of applications. Furthermore, its emergent connection with the field of quantum optics has revealed its potential for generating non-classical states of light. Here, we investigate the process of high-harmonic generation in semiconductors under a quantum optical perspective while using a Bloch-based solid-state description. Through the implementation of quantum operations based on the measurement of high-order harmonics, we demonstrate the generation of non-classical light states similar to those found when driving atomic systems. These states are characterized using diverse quantum optical observables and quantum information measures, showing the influence of electron dynamics on their properties. Additionally, we analyze the dependence of their features on solid characteristics such as the dephasing time and crystal orientation, while also assessing their sensitivity to changes in driving field strength. This study provides insights into HHG in semiconductors and its potential for generating non-classical light sources.
	\end{abstract}
	\maketitle
	
\section{INTRODUCTION}
Since the first observations of the high-harmonic generation process (HHG), in the last quarter of the 20th century~\cite{burnett_harmonic_1977,mcpherson_studies_1987,ferray_multiple-harmonic_1988} strong-field physics has undergone tremendous progress~\cite{corkum_attosecond_2007,krausz_attosecond_2009}. Nowadays, HHG stands as one of the primary mechanisms behind extreme ultraviolet (XUV) radiation sources~\cite{drescher_x-ray_2001,silva_spatiotemporal_2015} and attosecond pulse generation~\cite{krause_high-order_1992,popmintchev_bright_2012}. It has found diverse applications in non-linear XUV optics~\cite{kobayashi_27-fs_1998,midorikawa_xuv_2008,tsatrafyllis_ion_2016,chatziathanasiou_generation_2017,bergues_tabletop_2018,nayak_multiple_2018,senfftleben_highly_2020,orfanos_non-linear_2020}, attosecond science~\cite{krausz_attosecond_2009,amini_symphony_2019}, molecular tomography~\cite{itatani_tomographic_2004} and high-resolution spectroscopy~\cite{gohle_frequency_2005,cingoz_direct_2012,silva_high-harmonic_2018,alcala_high_2022}. 

In HHG, an input driving laser field within the infrared (IR) regime interacts with a target, resulting in the emission of coherent (although could also be incoherent~\cite{stammer_on_2023}) radiation with wavelengths significantly shorter than that of the incoming driving field. A hallmark of HHG spectra is a plateau of frequencies spanning from the infrared up to the XUV regime, followed by a sharp decay at a given cutoff frequency. The specific features of the generated radiation are profoundly influenced by target properties, leading to distinct characteristics depending on whether the radiation is generated from atomic~\cite{amini_symphony_2019,lhuillier_high-order_1993}, molecular~\cite{lynga_high-order_1996}, solid-state systems~\cite{ghimire_observation_2011,park_recent_2022,goulielmakis_high_2022} or nanostructures~\cite{ciappina_attosecond_2017}. Here, our focus centers on the HHG process within semiconductor materials. Unlike the conventional case of atoms, HHG in bulk matter show a linear increase of the cutoff frequency with the field's strength~\cite{ghimire_observation_2011,goulielmakis_high_2022} (in contrast to the quadratic trend in atoms~\cite{krause_high-order_1992}), a reduced dependence of the harmonic yield on the field's ellipticity~\cite{ghimire_observation_2011,goulielmakis_high_2022,corkum_plasma_1993,budil_influence_1993} and a strong-dependence on the crystal orientation~\cite{you_anisotropic_2017,kim_generation_2017}. Furthermore, the advantage of performing HHG in small ($\sim $500 $\mu$m) semiconductor samples~\cite{ghimire_observation_2011,schubert_sub-cycle_2014}, promises to extend the aforementioned applications to unprecedented scales.

High-harmonic generation in solid-state systems has also demonstrated significant potential towards quantum technology applications~\cite{lewenstein_attosecond_2022,bhattacharya_stronglaserfield_2023}. Research has revealed that driving HHG in strongly correlated materials can serve as a tool for detecting topological order~\cite{bauer_high-harmonic_2018,jurs_high-harmonic_2019,silva_topological_2019,jimenez-galan_lightwave_2020,chacon_circular_2020,baldelli_detecting_2022,pattanayak_role_2022} and for witnessing phase transitions~\cite{silva_high-harmonic_2018,roy_nonlinear_2020,orthodoxou_high_2021,kanega_linear_2021,murakami_high-harmonic_2021,murakami_anomalous_2022,uchida_high-order_2022}. In the context of the latter application, HHG was used to distinguish between different phases in high-$T_c$ superconductors~\cite{alcala_high_2022}, which represent one of the main platforms used for developing quantum information science applications~\cite{huang_superconducting_2020,RipollBook}. Concerning the direct use of semiconductors in the context of quantum information science, one of its main usages is as non-classical light sources~\cite{buckley_engineered_2012,uppu_scalable_2020,ollivier_reproducibility_2020,tomm_bright_2021,gonzalez-ruiz_violation_2022,gonzalez-ruiz_device_2022,coste_high-rate_2023}. In this direction, the emergent connection between quantum optics and strong-laser field physics~\cite{gorlach_quantum-optical_2020} has shown that, by means of HHG processes in atomic systems, one can generate non-classical states of light~\cite{lewenstein_generation_2021,rivera-dean_strong_2022,stammer_quantum_2023,bhattacharya_stronglaserfield_2023,pizzi_light_2023} with intensities strong enough to drive nonlinear processes in matter~\cite{lamprou_nonlinear_2023}, as well as massive frequency-entangled states~\cite{stammer_high_2022,stammer_theory_2022,stammer_iwo_2023}. Notably, these findings extend beyond HHG processes~\cite{rivera-dean_light-matter_2022} in atomic systems~\cite{rivera-dean_quantum_2023}. Specifically, for the case of solids, recent studies have demonstrated that the effect of electronic acceleration within the valence band could potentially yield analogous outcomes~\cite{gonoskov_nonclassical_2022} as those in Refs.~\cite{lewenstein_generation_2021,rivera-dean_strong_2022,stammer_quantum_2023,stammer_high_2022}. Furthermore, studies under a Wannier-Bloch description highlighted the possibility of having entanglement between the electronic and field degrees of freedom~\cite{rivera-dean_entanglement_2023}.

While these studies in solid-state systems have provided essential insights about the generation of non-classical states of light after HHG in this kind of media, they are either limited to specific electron dynamics within the solid~\cite{gonoskov_nonclassical_2022}, or involve simplifications on the interaction with the material, restricting the analysis to a limited number of Wannier sites and neglecting dephasing effects~\cite{rivera-dean_entanglement_2023}. These limitations can be addressed by employing a Bloch-based description of the solid system, where the semiconductor Bloch equations (SBEs)~\cite{chacon_circular_2020,yue_introduction_2022,li_high_2023} offer a straightforward framework for describing dephasing times and the incorporation of many-body effects. 

In this study, we explore the utilization of HHG in semiconductor materials to generate non-classical light states using a Bloch-based description of the solid system, alongside a quantum mechanical description of the electromagnetic field. Specifically, we focus on ZnO and analyze its interaction with a linearly polarized field aligned along different crystal directions, examining how varying dephasing times and field strengths affect the final results. By introducing conditioning operations into the HHG process, analogous to those described in Refs.~\cite{lewenstein_generation_2021,rivera-dean_strong_2022,stammer_quantum_2023,stammer_high_2022}, we demonstrate the generation of non-classical states of light. These states are characterized using various quantum optical observables and quantum information measures, highlighting the influence of the different electron dynamics on their features. Specifically, we observe two non-exclusive indications of non-classical behaviors: the presence of Wigner function negativities and non-negligible entanglement between the electromagnetic field modes, each of which is observed under two distinct regimes.

The text is organized as follows. In Section~\ref{Sec:Theory} we provide the theoretical background, beginning with a semiclassical description of HHG in semiconductors, followed by our quantum optical analysis. Section~\ref{Sec:Results}, summarizes the main findings of the paper, where we compute the Wigner function of the generated light states and assesses other quantum information measures such as the fidelity and linear entropy. Finally, our conclusions are presented in Section~\ref{Sec:Conclusions}, where we discuss our results and outline potential avenues for future research.

\section{THEORETICAL DESCRIPTION}\label{Sec:Theory}
The theoretical analysis presented in this work considers excitations over a ZnO material driven by an intense linearly polarized laser field along various crystal orientations. Specifically, we focus on polarizations along the $\Gamma - M$ and $\Gamma-A$ directions, which we will subsequently denote as the $x$ and $z$ orientations, respectively. Our ZnO modeling employs a two-band, tight-binding model, ensuring that the subsequent band-dispersion relations hold for the valence ($v$) and conduction ($c$) bands
\begin{align}
	& E_{v}(\vb{k})
		= \sum_{i=x,y,z}
			\sum_{j=0}^{\infty}
			\alpha^j_{v,i}
				\cos(j k_{i} a_{i}),\label{Eq:val:dis}
	\\&E_{c}(\vb{k})
	= E_g +  \sum_{i=x,y,z}	
	\sum_{j=0}^{\infty}
	\alpha^j_{c,i}
	\cos(j k_{i} a_{i}),\label{Eq:cond:dis}
\end{align}
where $a_{i}$ represents the lattice constant along the $i$ direction, $E_g$ stands for the band-gap energy at the $\Gamma$ point, and $\alpha_{m,i}^{(j)}$ denotes the expansion coefficients for band $m$ in the $i$ direction. These parameters are chosen in accordance with the seminal works by G.~Vampa et al~\cite{vampa_theoretical_2014,vampa_semiclassical_2015,vampa_high-harmonic_2015}, and their specific values are provided in Appendix~\ref{App:Num:Params}. However, in contrast to these works, we do account for the $k$-dependence of the transition matrix elements between the conduction and valence bands, denoted as $d^{(i)}_{vc}(\vb{k}) = \sqrt{E_{p,i}/[2(E_c(\vb{k})-E_v(\vb{k}))^2]}$ (see e.g. Ref.~\cite{HaugBookch5}), where $E_p$ corresponds to the Kane parameter, varying with the particular band under consideration. For ZnO, these parameters are given by $E_{p,x} = E_{p,y} = 0.355$ a.u. and $E_{p,z} = 0.479$ a.u., as indicated in Refs.~\cite{vurgaftman_band_2001,ozgur_comprehensive_2005,schleife_band-structure_2009,yan_strain_2012}. 

\subsection{Semiclassical description}
In this subsection, we provide a summary of the equations that delineate the HHG process in solid-state systems within a semiclassical framework. Subsequently, we build our quantum optical framework upon this foundation. Thus, within the length gauge, and under the single-active electron and dipole approximations, the Hamiltonian that characterizes the interplay between the crystal and a (classical) intense laser field is expressed as
\begin{equation}
	\hat{H}_{\text{sc}}(t)
		= \hat{H}_{\text{cr}}
			+ \mathsf{e} \hat{r}_{i} E_{\text{cl}}(t),
\end{equation} 
where $H_{\text{cr}}$ represents the crystal Hamiltonian, modeled as previously specified, and $E_{\text{cl}}(t)$ characterizes the classical input, linearly polarized, driving electric field, with $\mathsf{e}$ the electronic charge. In the following, we divide the position operator $\hat{r}_i$ into its interband and intraband components, whose matrix elements in the Bloch basis $\{\ket{\phi_{\vb{k},m}}\}$, with $\vb{k}$ denoting the crystal momentum and $m$ designating the associated band, are given by (see e.g. Ref.~\cite{yue_introduction_2022})
\begin{align}
	&\mel{\phi_{\vb{k},m}}{\hat{r}_{i,\text{tra}}}{\phi_{\vb{k}',l}}
		= i\hbar \delta_{m,l} \pdv{}{k_i} \delta(\vb{k}-\vb{k}'),\label{Eq:intra:els}
	\\
	&\mel{\phi_{\vb{k},m}}{\hat{r}_{i,\text{ter}}}{\phi_{\vb{k}',l}}
		= \mathsf{e}^{-1}d^{(i)}_{ml} (\vb{k}) \delta(\vb{k}-\vb{k}').\label{Eq:inter:els}
\end{align}

Thus, by considering this division, the Hamiltonian reads as
\begin{equation}\label{Eq:H:sc}
		\hat{H}_{\text{sc}}(t)
	= \hat{H}_{\text{cr}}
		+ \mathsf{e} \hat{r}_{i,\text{tra}} E_{\text{cl}}(t)
		+ \mathsf{e} \hat{r}_{i,\text{ter}} E_{\text{cl}}(t),
\end{equation}
such that the Schrödinger equation describing the dynamics of the electronic system is given by
\begin{equation}\label{Eq:sc:Sch:without}
	i \hbar \pdv{\ket{\psi(t)}}{t}
		= 	\hat{H}_{\text{sc}}(t) \ket{\psi(t)}.
\end{equation}

From the definitions in Eqs.~\eqref{Eq:intra:els} and Eq.~\eqref{Eq:inter:els}, one can see that in the crystal momentum representation this equation couples different values of $k$, which runs over a continuum range of values. With the aim of simplifying its analysis, we first introduce the unitary transformation $\ket{\psi(t)} = e^{i \mathsf{e}A_{\text{cl}}(t) \hat{r}_{i,\text{tra}}/\hbar} \ket{\psi'(t)}$, where $A_{\text{cl}}(t)$ is the classical vector potential, defined as $E_{\text{cl}}(t) \equiv - \partial \! A_{\text{cl}}/\partial t$. Through this transformation, our Schrödinger equation gets modified to
\begin{equation}\label{Eq:sc:Sch:with}
		\begin{aligned}
		i \hbar \pdv{\ket{\psi'(t)}}{t}
			= 
			\big[ 
				&
					\hat{H}_{\text{cr}}(t)
					+
					\mathsf{e}\hat{r}_{i,\text{ter}}(t)E_{\text{cl}}(t)
			\big]
			\ket{\psi'(t)},
		\end{aligned}
\end{equation}
where we have defined
\begin{align}
	 &\hat{H}_{\text{cr}}(t) 
	 	\equiv  e^{-i \mathsf{e}A_{\text{cl}}(t) \hat{r}_{i,\text{tra}}/\hbar}  \hat{H}_{\text{cr}} e^{i \mathsf{e}A_{\text{cl}}(t) \hat{r}_{i,\text{tra}}/\hbar},
	 	\label{Eq:Hcr:transf}
	 \\
	& \hat{r}_{i,\text{ter}}(t)
		\equiv 
		e^{-i \mathsf{e}A_{\text{cl}}(t) \hat{r}_{i,\text{tra}}/\hbar}
			\hat{r}_{i,\text{ter}}
		e^{i \mathsf{e}A_{\text{cl}}(t) \hat{r}_{i,\text{tra}}/\hbar}.
		\label{Eq:rter:transf}
\end{align}

This transformation shifts our frame of reference to that of the oscillating electron within the field, effectively allowing us to operate within the frame of the canonical crystal momentum $\vb{K}$. This momentum is defined as $\vb{k} = \vb{K} +\mathsf{e}\boldsymbol{\mathcal{E}}_i A_{\text{cl}}(t)$ where $\boldsymbol{\mathcal{E}}_i$ denotes the polarization direction, and remains a constant of motion. Consequently, upon projecting our equations relative to a given $\ket{\vb{K},m}$, with $m$ denoting the band, we obtain
\begin{equation}\label{Eq:scl:Sch}
	\begin{aligned}
	i \hbar \pdv{b_m(\vb{K},t)}{t}
		&= E_m\big(\vb{K} + \mathsf{e}\boldsymbol{\mathcal{E}}_i A_{\text{cl}}(t)\big)
			b_m(\vb{K},t)
			\\ & \quad
			+ \sum_{l=c,v} d_{ml}^{(i)}
					\big(\vb{K} + \mathsf{e}\boldsymbol{\mathcal{E}}_i A_{\text{cl}}(t)\big)
						b_l(\vb{K},t)
	\end{aligned}
\end{equation}

Hence, the main advantage of working within this shifted frame of reference is that the resulting differential equations become uncoupled in relation to $\vb{K}$. In the asymptotic limit $t\to \infty$, corresponding to the end of the pulse, it coincides with the electron's final crystal momentum. We note that, within our description, the Berry connection terms $d^{(i)}_{mm}(\vb{k})$ are incorporated. However, for centrosymmetric materials, these terms vanish, as it happens with ZnO when the driving laser field is linearly polarized along the $\Gamma-M$ direction~\cite{jiang_role_2018}.

From the system of differential equations defined in Eq.~\eqref{Eq:scl:Sch}, one can derive the semiconductor Bloch equations (SBEs)~\cite{chacon_circular_2020,yue_introduction_2022,li_high_2023} that delineate the dynamics of the bands' population, denoted as $n_{m}(\vb{K},t) \equiv b_m(\vb{K},t)^*b_m(\vb{K},t)$, and the coherences between them, represented as $\pi(\vb{K},t) = b_v(\vb{K},t)^*b_c(\vb{K},t)$. These equations are given as follows
\begin{align}
	&i\hbar \pdv{n_m(\vb{K},t)}{t}
		= \theta_m
			E_{\text{cl}}(t) [d_{cv}^{(i)}\big(\vb{K} + \mathsf{e}\boldsymbol{\mathcal{E}}_i A_{\text{cl}}(t)\big)]^*
			\pi(\vb{K},t) \nonumber
			\\&\hspace{2.5cm}+ \text{c.c.}\label{Eq:semic:pop}
	\\
	&i\hbar \pdv{\pi(\vb{K},t)}{t}
		= \Big[
				\mathcal{E}_g\big(\vb{K}+\mathsf{e}\boldsymbol{\mathcal{E}}_iA_{\text{cl}}(t)\big)\nonumber
				\\
				&\hspace{1.9cm}+ E_{\text{cl}}(t) \xi^{(i)}_g\big(\vb{K}+\mathsf{e}\boldsymbol{\mathcal{E}}_iA_{\text{cl}}(t)\big)
				- i \dfrac{1}{T_2}
			\Big]\pi(\vb{K},t)\nonumber
			\\&\hspace{1.9cm}
			+E_{\text{cl}}(t)
				d^{(i)}_{cv}\big(\vb{K}+\mathcal{E}_iA_{\text{cl}}(t)\big)
				w(\vb{K},t),\label{Eq:semic:coh}
\end{align}
where we have defined $\theta_v \equiv 1$, $\theta_c \equiv -1$, the bandgap energy as $\mathcal{E}_g(\vb{k}) \equiv E_c(\vb{k}) - E_v(\vb{k})$, the difference between the Berry connections of each band as $\xi^{(i)}_g(\vb{k}) \equiv d^{(i)}_{cc}(\vb{k}) - d^{(i)}_{vv}(\vb{k})$ and the population difference between bands as $w(\vb{K},t) \equiv n_v(\vb{K},t)-n_c(\vb{K},t)$. Additionally, we incorporate a phenomenological description of electron-electron and electron-phonon couplings through the dephasing time $T_2$ into our equations. In the following, we will see how this parameter influences the HHG spectrum.

The mechanism of HHG in semiconductors is typically attributed to two distinct dynamics: intra- and interband processes~\cite{golde_high_2008,golde_microscopic_2009,ghimire_observation_2011,ghimire_generation_2012,schubert_sub-cycle_2014,vampa_theoretical_2014}. In the former, the emission of radiation is often linked to electronic scattering within the non-parabolic energy dispersion profile of a specific band, commonly referred to as Bloch oscillations~\cite{bloch_uber_1929}. In contrast, interband dynamics follow the three-step-like model of the HHG mechanism, where: (1) an electron-hole pair is created through the promotion of an electron from the valence to the conduction band, (2) the pair gets accelerated by the field within their respective bands, and (3) the electron-hole recombination occurs within the valence band, leading to the emission of a photon whose frequency depends on the energy difference between the bands~\cite{vampa_theoretical_2014,vampa_semiclassical_2015}. The respective current components giving rise to these effects are
\begin{align}
	&j_{i,\text{tra}}
		=-\mathsf{e} \tr(\comm{\hat{r}_{i,\text{tra}}}{\hat{H}_{\text{cr}}}\hat{\rho}(t))
		\label{Eq:intra:curr},
	\\
	&j_{i,\text{ter}}
		= \mathsf{e} \dv{}{t}
		\big[ 
			\tr(\hat{r}_{i,\text{ter}}\hat{\rho}(t))
		\big]
		\label{Eq:inter:curr},
\end{align}
where $\hat{\rho}(t)$ is a density matrix with populations and coherences given by $n_m(\vb{K},t)$ and $\pi(\vb{K},t)$, respectively, when expanding $\hat{\rho}(t)$ in the canonical crystal momentum basis.

\begin{figure*}
	\centering
	\includegraphics[width = 0.7\textwidth]{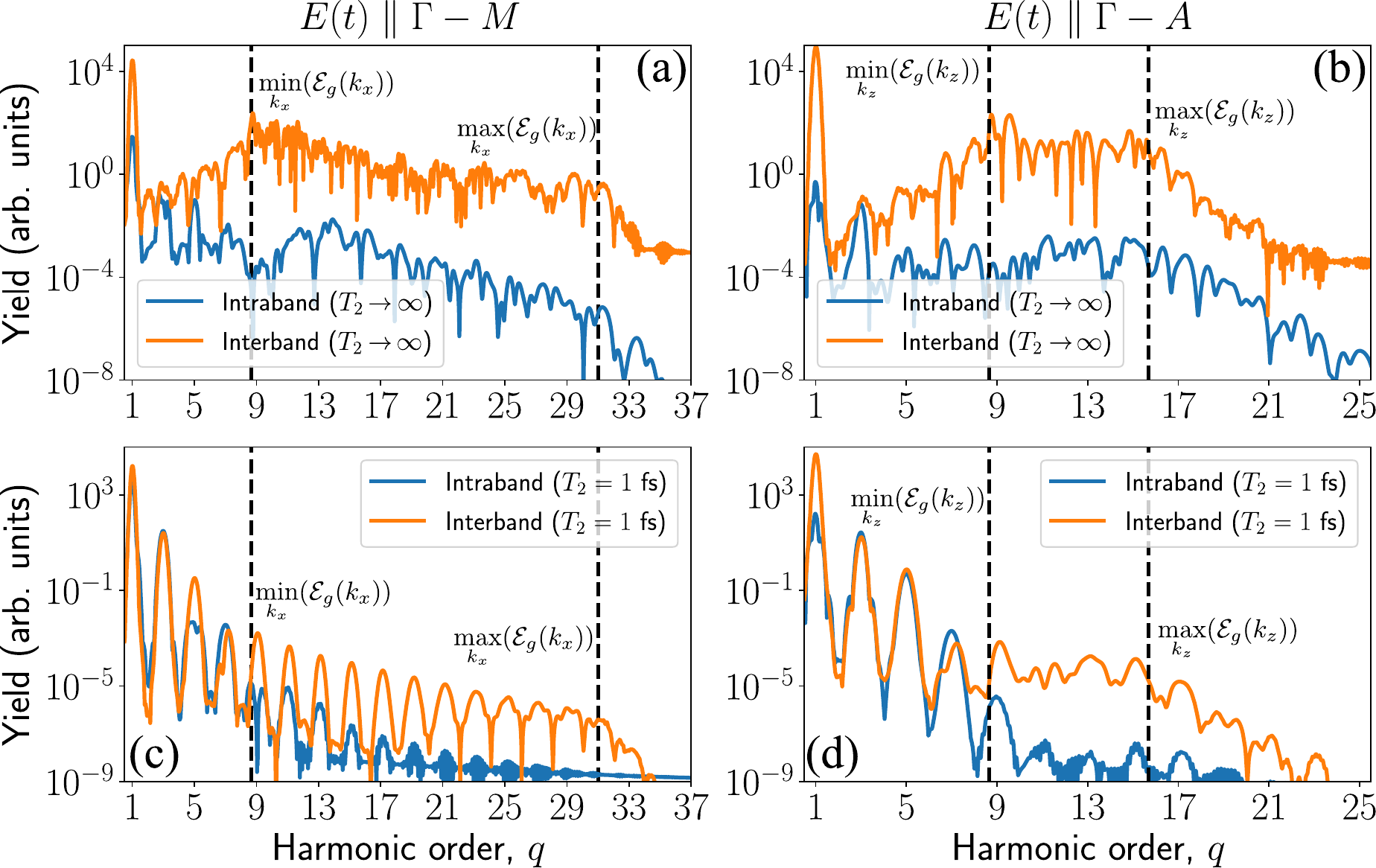}
	\caption{High-harmonic generation spectra for laser fields polarized linearly along distinct solid directions. The laser field features a Gaussian envelope with a central wavelength $\lambda_L = 3.25$ $\mu$m, field strength of 0.5 V/$\angstrom$ and a duration of $\Delta t \sim 96$ fs (equivalent to 9 optical cycles). Specifically, for panels (a) and (c), we examine the case of the $\Gamma-M$ direction, while for panels (b) and (d), we consider the case of the $\Gamma-A$ direction. In panels (a) and (b), infinite dephasing times are assumed, while in panels (c) and (d), a finite $T_2 = 1$ fs is employed, resulting in well-resolved harmonic peaks.}
	\label{Fig:HHG:semiclassic:spec}
\end{figure*}

While these are generally regarded as the primary contributors to HHG emission in semiconductors, the mechanisms underpinning this process have undergone extensive scrutiny in the literature~\cite{yue_introduction_2022}. In the context of non-centrosymmetric materials, where Berry connection terms are non-zero, the anomalous current assumes a pivotal role~\cite{liu_high-harmonic_2017,banks_dynamical_2017,luu_measurement_2018,avetissian_high_2020,chacon_circular_2020}, giving rise to the imperfect recollision mechanism in solid-state HHG~\cite{yue_imperfect_2020,li_high_2023}. Another contributing factor is commonly known as the mixture current, which arises due to the interplay between inter- and intraband operators~\cite{wilhelm_semiconductor_2021}, i.e. $[\hat{r}_{i,\text{tra}},\hat{r}_{i,\text{ter}}]$, although its implications have remained largely unexplored~\cite{yue_introduction_2022}. Consequently, our analysis of ZnO predominantly centers on the effects that the mechanisms outlined in Eqs.~\eqref{Eq:intra:curr} and \eqref{Eq:inter:curr} exert on the quantum optical state after the HHG process.

The influence of each current contributions presented in Eqs.~\eqref{Eq:intra:curr} and \eqref{Eq:inter:curr} on the HHG spectrum, computed as $\omega^2|\text{FT}[j_{i,\text{ter}}+j_{i,\text{tra}}]|^2$, is illustrated in Fig.~\ref{Fig:HHG:semiclassic:spec}. The spectrum is calculated for laser fields polarized along the $\Gamma-M$ (Figs.~\ref{Fig:HHG:semiclassic:spec}~(a) and (c)) and the $\Gamma-A$ (Figs.~\ref{Fig:HHG:semiclassic:spec}~(b) and (d)) directions~\footnote{In our numerics, we have set the transition dipole matrix elements to be real numbers, in the spirit of the works of Vampa et al \cite{vampa_theoretical_2014,vampa_semiclassical_2015}. While this is valid for centrosymmetric excitations, as it happens when the field is polarized along the $\Gamma-M$ crystal direction of ZnO, this is not true in general and specifically when the the field is polarized along the $\Gamma-A$ direction \cite{jiang_role_2018}. In fact, the complex part of the transition dipole elements becomes crucial here as it leads to the presence of even harmonic orders \cite{jiang_effect_2017}. This indeed happens when the laser field is linearly polarized along the $\Gamma-A$ ZnO crystal direction, as shown in \cite{jiang_role_2018}. However, we do not expect these contributions to play a fundamental role when computing the different quantum optical observables and the different quantum information measures.}. We observe that, in both cases, the interband contribution (shown in orange) predominates over the intraband within the non-perturbative regime. This regime is delineated between the minimum and maximum bandgap energies for each solid direction (vertical dashed lines). However, in the perturbative region of the spectrum (below $\min_{\vb{k}}(\mathcal{E}_g(\vb{k}))$) a clear dominance of one contribution over the other is not apparent. Furthermore, it is noteworthy that incorporating finite dephasing times yields to resolved harmonic peaks when $T_2 \sim 1$ fs~\cite{vampa_theoretical_2014,vampa_semiclassical_2015} (Figs.~\ref{Fig:HHG:semiclassic:spec}~(c) and (d)), along with reduced harmonic conversion efficiencies. While the use of such small dephasing times has been common practice in these approaches, it has remained a topic of controversy within the community~\cite{goulielmakis_high_2022}. In this context, various frameworks for describing the laser-solid interaction can yield distinct insights. For instance, in Ref.~\cite{brown_real-space_2022} the need of extremely short dephasing times was circumvented by adopting a Wannier-based approach, where long spatial trajectories interefere destructively.

\subsection{Quantum optical description}\label{Sec:QO:des}
In this section, we expand upon the previously outlined description into the quantum optical domain. This involves substituting the classical electric field $E_{\text{cl}}(t)$ in Eq.~\eqref{Eq:H:sc} with the electric field operator $\hat{E}(t)$ and incorporating the free-field Hamiltonian $\hat{H}_{\text{field}}$. Consequently, the Hamiltonian that characterizes the interplay between the crystal and the field, for a single-active electron within the dipole approximation and under the length gauge, can be expressed as
\begin{equation}
	\hat{H}(t)
		= \hat{H}_{\text{cr}} 
			+ \mathsf{e} \hat{r}_{i,\text{tra}} \hat{E}(t)
			+ \mathsf{e} \hat{r}_{i,\text{ter}} \hat{E}(t)
			+ \hat{H}_\text{field}.
\end{equation}

While our objective is to capture laser-matter interactions involving electromagnetic fields of finite duration, which inherently necessitate encompassing the complete continuum spectrum, for simplicity we focus on a discrete set of modes. This discrete set spans from the central frequency $\omega_L$ of the driving field, to the cutoff region of the harmonic spectrum, defined as $\omega_{q_c} = q_c \omega_L$, i.e., ${\omega_q = q \omega_L : q=1,2,\cdots, q_c}$. Consequently, we formulate the free-field Hamiltonian for linearly polarized fields as $\hat{H}_{\text{field}} = \sum_{q=1}^{q_{\text{cutoff}}} \hbar \omega_q \hat{a}^\dagger_q \hat{a}_q$, where $\hat{a}_q$ ($\hat{a}^\dagger_q$) represents the annihilation (creation) operator for the field mode with frequency $q$. Similarly, we characterize the electric field operator as
\begin{equation}\label{Eq:E:def}
	\hat{\vb{E}}(t)
		= -if(t) \sum^{q_c}_{q=1} \vb{g}(\omega_q) 
						\big( 
							\hat{a}_q^\dagger 
							- \hat{a}_q
						\big),
\end{equation}
where $0\leq f(t) \leq 1$ is a dimensionless function accounting for the laser field envelope. Here, $\vb{g}(\omega_q) \equiv \boldsymbol{\mathcal{E}}_i\sqrt{\hbar \omega_q/(2\epsilon_0 V)}$ is a factor arising from the expansion of the electric field operator into the field modes~\cite{ScullyBookCh1,Gerry__Book_2005_ch2,VogelBookch2} with $\boldsymbol{\mathcal{E}}_i$ a unitary vector pointing in the direction along which the field is polarized, $\epsilon_0$ the vacuum permittivity and $V$ the quantization volume. Hereafter, we denote $\hat{E}(t)$ as the electric field operator component along the $i$ direction, i.e. $\hat{E}(t)\equiv \boldsymbol{\varepsilon}_i \cdot \hat{\vb{E}}(t)$, which aligns with the polarization of the input driving field.

In this context, the initial state of the joint system is given by
\begin{equation}
	\ket{\Psi(t=t_0)}
		= \ket{\phi_{\vb{k}_0,v}} \otimes \ket{\alpha_L} \bigotimes_{q=2}^{q_c}\ket{0_q},
\end{equation}
where the electron is initially located in the valence band with crystal momentum $\vb{k}_0$. Concerning the quantum optical state, the infrared driving mode is in a coherent state of amplitude $\alpha_L$, while all the other modes are in a vacuum state. Furthermore, the dynamics of the composite system are described by the Schrödinger equation
\begin{equation}
	i \hbar \pdv{\ket{\Psi(t)}}{t}
		= \hat{H}(t) \ket{\Psi(t)},
\end{equation}
which in the interaction picture with respect to $\hat{H}_{\text{field}}$ reads
\begin{equation}\label{Eq:H:QO:I}
	i \hbar \pdv{\ket{\bar{\Psi}(t)}}{t}
		= \big[
				\hat{H}_{\text{sc}}(t)
				+ \mathsf{e} \hat{r}_{i,\text{tra}}\hat{E}(t)
				+ \mathsf{e} \hat{r}_{i,\text{ter}} \hat{E}(t)
			\big]\ket{\bar{\Psi}(t)},
\end{equation}
where the electric field operator has gained an additional time dependence, that is,
\begin{equation}
	\hat{\vb{E}}(t)
	= -if(t) \sum^{q_c}_{q=1} \vb{g}(\omega_q) 
	\big( 
	\hat{a}_q^\dagger e^{i\omega_q t}
	- \hat{a}_q e^{-i\omega_q t}
	\big).
\end{equation}

In Eq.~\eqref{Eq:H:QO:I}, $\ket{\bar{\Psi}(t)} = e^{-i\hat{H}_{\text{field}} t/\hbar} \hat{D}_1(\alpha_L) \ket{\Psi(t)}$, with $\hat{D}_q(\alpha) \equiv \exp[\hat{a}^\dagger_q \alpha^* - \hat{a}_q\alpha]$ the displacement operator acting on the electromagnetic field mode $q$~\cite{ScullyBookCh2,Gerry__Book_2005_ch3,VogelBookch3}. Additionally, $\hat{H}_{\text{sc}}(t)$ corresponds to the semiclassical Hamiltonian presented in Eq.~\eqref{Eq:scl:Sch}, where $E_{\text{cl}}(t) = \text{tr}(\hat{E}(t) \dyad{\Psi(t_0)})$.

Following a similar approach to the one employed when transitioning from Eq.~\eqref{Eq:sc:Sch:without} to Eq.~\eqref{Eq:sc:Sch:with}, we now introduce the transformation
\begin{equation}
	\ket{\bar{\Psi}(t)} =
		e^{i\mathsf{e}(A_{\text{cl}}(t) + \hat{A}(t))\hat{r}_{i,\text{tra}}/\hbar}
			 \ket{\bar{\Psi}'(t)},
\end{equation}
where $\hat{A}(t)$ is the vector potential, which is defined as $\hat{E}(t) \equiv -\partial\!\hat{A}(t)/\partial t$. Thus, incorporating this definition into Eq.~\eqref{Eq:H:QO:I} and considering terms up to first order with respect to $\text{g}(\omega_q)$ (see Appendix \ref{Appendix:A:I}), we arrive at
\begin{equation}\label{Eq:Sch:QO:1}
	\begin{aligned}
	i \hbar \pdv{\ket{\bar{\Psi}'(t)}}{t}
		&\approx 
		\Big[
			\hat{H}_{\text{cr}}(t)
			+ \mathsf{e}\hat{r}_{i,\text{ter}}(t)
				\big(
					E_{\text{cl}}(t)
					+ \hat{E}(t)
				\big)
				\\&\hspace{1.6cm}
				- i \mathsf{e}\hbar^{-1}
					\hat{v}_{i,\text{tra}}(t)
						\hat{A}(t)
			\Big]
			\ket{\bar{\Psi}'(t)},
	\end{aligned}
\end{equation}
where we additionally disregarded the influence of the mixture current. Here, we defined $\hat{r}_{i,\text{ter}}(t)$ as in Eq.~\eqref{Eq:rter:transf}, and $\hat{v}_{i,\text{tra}} \equiv [\hat{r}_{i,\text{tra}},\hat{H}_{\text{cr}}]$, such that $\hat{v}_{i,\text{tra}}(t)$ is obtained as in Eq.~\eqref{Eq:rter:transf} upon the change $\hat{r}_{i,\text{ter}}\to \hat{v}_{i,\text{tra}}$.

The first two terms on the right-hand side of Eq.~\eqref{Eq:Sch:QO:1} contribute to the Hamiltonian that leads to the semiclassical evolution described by Eq.~\eqref{Eq:sc:Sch:with}. Hence, the last transformation we introduce involves moving to the interaction picture with respect to that term, meaning $\ket{\bar{\Psi}'(t)} = \hat{U}_{\text{sc}}(t,t_0)\tPsi$, where $\hat{U}_{\text{sc}}(t,t_0) \equiv \hat{\mathcal{T}}\text{exp}[-i \int^t_{t_0}\dd \tau \hat{\bar{H}}_{\text{sc}}(\tau)/\hbar]$ with $\hat{\mathcal{T}}$ being the time-ordering operator and $\hat{\bar{H}}_{\text{sc}}(t) \equiv \hat{H}_{\text{cr}}(t) + \mathsf{e}\hat{r}_{i,\text{ter}}E_{\text{cl}}(t)$. Subsequently, we obtain
\begin{equation}
	i\hbar \pdv{\tPsi}{t}
		= \Big[
				\mathsf{e}\hat{\bar{r}}_{i,\text{ter}}(t) \hat{E}(t)
				- i \mathsf{e} \hbar^{-1}\hat{\bar{v}}_{i,\text{tra}}\hat{A}(t)
			\Big]\tPsi,
\end{equation}
where $\hat{\bar{r}}_{i,\text{ter}}(t) \equiv \hat{U}^\dagger_{\text{sc}}(t,t_0)\hat{r}_{i,\text{ter}}(t)\hat{U}_{\text{sc}}(t,t_0)$ and $\hat{\bar{v}}_{i,\text{tra}}(t) \equiv \hat{U}^\dagger_{\text{sc}}(t,t_0)\hat{v}_{i,\text{tra}}(t)\hat{U}_{\text{sc}}(t,t_0)$.

A solution to this differential equation, for a two-band model, can always be written as
\begin{equation}
	\tPsi
		= \sum_{m=v,c}
				\int \dd\vb{K}
					\ket{\vb{K},m}\ket{\Phi_m(\vb{K},t)},
\end{equation}
where $\ket{\Phi_{m}(\vb{K,t})} = \langle \vb{K},m\tPsi$ represents the quantum optical state when the electron resides in band $m$ with canonical crystal momentum $\vb{K}$. By considering terms up to first order in $\text{g}(\omega_L)$ and accounting for the initial conditions, we derive the subsequent expression for the quantum optical component (see Appendix~\ref{Appendix:A:II} for a detailed derivation)
\begin{align}
	&\ket{\Phi_{v}(\vb{K},t)}
	\approx \delta(\vb{K}-\vb{K}_0)
			\hat{\mathcal{D}}
			\big(
				\boldsymbol{\chi}_{v}(\vb{K},t,t_0)
			\big)
			\bigotimes^{q_c}_{q=1}\ket{0_q},\label{Eq:QO:state:val}
	\\
	& \ket{\Phi_{c}(\vb{K},t)}
		\approx -	\dfrac{i}{\hbar}\nonumber
	 	\delta(\vb{K}-\vb{K}_0)
	 	\\ &\hspace{1.5cm}\times 
		\int^t_{t_0} \dd t'
			\hat{\mathcal{D}}
			\big(
				\boldsymbol{\chi}_{c}(\vb{K},t,t')
			\big)
			\hat{\mathcal{M}}_{c,v}(\vb{K},t')\nonumber
		\\&\hspace{2.75cm}\times
		\hat{\mathcal{D}}
			\big(
				\boldsymbol{\chi}_{v}(\vb{K},t',t_0)
			\big)
			\bigotimes^{q_c}_{q=1}\ket{0_q}
			\label{Eq:QO:state:cond},
\end{align}
where $\vb{K}_0$ represents the initial crystal momentum of the electron. In these two expressions, we define $\hat{\mathcal{D}}(\boldsymbol{\chi}_{\mathsf{i}}(\vb{K},t,t_0)) \equiv \prod^{q_c}_{q=1}[e^{i\varphi_q(\vb{K},t,t_0)}\hat{D}_{q}(\chi^{(q)}_{\mathsf{i}}(\vb{K},t,t_0))]$, with $\varphi_q(\vb{K},t,t_0)$ as a phase factor arising from the use of the Baker–Campbell–Hausdorff (BCH) formula~\cite{Gerry__Book_2005_ch2,ScullyBookCh2,lewenstein_generation_2021,rivera-dean_strong_2022,stammer_quantum_2023} to express the solution above, and where $\chi^{(q)}_{\mathsf{i}}(\vb{K},t,t_0)$ is given by
\begin{equation}\label{Eq:displacements}
	\begin{aligned}
	\chi^{(q)}_{\mathsf{i}}\big(\vb{K},t,t_0\big)
		&= \dfrac{1}{\hbar}g(\omega_q)
			\int^t_{t_0} \dd \tau
			\Big[
				-M_{\mathsf{i},\mathsf{i}}^{(\text{ter})}(\vb{K},\tau)f_q(\tau)
				\\&\hspace{3cm}
				+M_{\mathsf{i},\mathsf{i}}^{(\text{tra})}(\vb{K},\tau)F_q(\tau)
			\Big],
	\end{aligned}
\end{equation}
where $f_q(t) \equiv f(t)e^{i\omega_q t}$, $F_q(t) \equiv \int \dd t f_q(t)$, and $\hat{M}^{(\text{ter})}_{\mathsf{i,j}}(\vb{K},t)$ and $\hat{M}^{(\text{tra})}_{\mathsf{i,j}}(\vb{K},t)$ account for the transition matrix elements of the operators $\hat{\bar{r}}_{i,\text{ter}}(t)$ and $\hat{\bar{v}}_{i,\text{tra}}$ between states in bands $\mathsf{i}$ and $\mathsf{j}$ (see Appendix \ref{Appendix:A:II}), respectively, with $\hat{\mathcal{M}}_{c,v}(\vb{K},t)$ in Eq.~\eqref{Eq:QO:state:cond} given by 
\begin{equation}
	\hat{\mathcal{M}}_{\mathsf{i},\mathsf{j}}(\vb{K},t')
	\equiv		
	M^{(\text{ter})}_{\mathsf{i},\mathsf{j}}(\vb{K},t)\hat{E}(t)
	+\hbar^{-1}M^{(\text{tra})}_{\mathsf{i},\mathsf{j}}\hat{A}(t).
\end{equation}

Similar to the quantum optical description of strong-field processes in atomic systems of Refs.~\cite{lewenstein_generation_2021,rivera-dean_strong_2022,stammer_high_2022,stammer_theory_2022,stammer_quantum_2023}, in solid-state, each electromagnetic field mode experiences a displacement by an amount $\chi^{(q)}_v(\vb{K},t,t_0)$. However, for solid-state systems, this quantity is influenced by both the interband and intraband dynamics of electrons. In the following, we aim to examine the impacts of these dynamics following HHG processes. To achieve this, we begin by reverting the transformations conducted to arrive at Eq.~\eqref{Eq:Sch:QO:1}. Yet, we continue to work within the shifted and rotating frame of reference for the quantum optical state, yielding
\begin{equation}\label{Eq:Back:to:Original}
	\ket{\bar{\Psi}(t)}
		=  e^{i\mathsf{e}(A_{\text{cl}}(t) + \hat{A}(t))\hat{r}_{i,\text{tra}}/\hbar}
			\hat{U}_{\text{sc}}(t,t_0)
			\toPsi.
\end{equation}

It is crucial to recognize at this point that, in Eq.~\eqref{Eq:Back:to:Original}, the unitary transformation $ e^{i\mathsf{e} \hat{A}(t))\hat{r}_{i,\text{tra}}/\hbar}$ introduces entanglement between the canonical crystal momentum and the field modes, given that $\hat{r}_{i,\text{\text{tra}}}$ is not diagonal with respect to $\vb{K}$. Nonetheless, in the discrete description we are using, where the field's envelope is accounted for using the weight function $f(t)$ in the definition of the electric field operator, as shown in Eq.~\eqref{Eq:E:def}, it becomes apparent that in the asymptotic limit $t\to \infty$ when the pulse concludes, $\hat{A}(t) \to 0$ (see Fig.~\ref{Fig:App:Envelopes} in Appendix \ref{Appendix:A:II}). In this regime, and for the discrete mode analysis we present here, we get
\begin{equation}
	\ket{\bar{\Psi}(t)}
	= 	\hat{U}_{\text{sc}}(t,t_0)
	\toPsi.
\end{equation}

As mentioned earlier, our focus lies in HHG, such that the electron ends up in the valence band of the solid system. Accordingly, we restrict to this process by applying the projector operator $\hat{P}_v = \int \dd \vb{K} \dyad{\vb{K},v}{\vb{K},v}$ to our state, and tracing out the electronic degrees of freedom. Hence, we arrive at the quantum optical state
\begin{equation}\label{Eq:final:QO:state}
	\ket{\Phi_v(\vb{K},t)}
		\approx \hat{\mathcal{D}}
				\big(
					\boldsymbol{\chi}_v(\vb{K}_0,t,t_0)
				\big)
				\bigotimes_{q=1}^{q_c}\ket{0_q},
\end{equation}
similarly to what is found for atomic systems. It is worth noting that, here, we have neglected contributions of the form $\langle \vb{K},v\vert U_{\text{sc}}(t,t_0)\vert \vb{K}',c\rangle$ which are smaller compared to the central one shown in the equation above (they scale as $\text{g}(\omega_L)$), but that introduce entanglement between the electron and field degrees of freedom (see Appendix \ref{Appendix:A:II}), as recently stated in Ref.~\cite{rivera-dean_entanglement_2023}. However, as shown in this study, the amount of entanglement observed in the case of ZnO was nearly negligible. 

It is imperative to bear in mind, up to this point, that our analysis has focused on the single-electron level. However, the ground state of a semiconductor consists of a fully occupied valence band, forming a Fermi sea where each electron initially possesses a well-defined crystal momentum $\vb{K}$ and interacts with other electrons. Notably, in the formulation of the SBEs, many-body interactions are accounted for through effective masses and couplings. Hence, within this framework, we can treat each electron independently encompassing various initial values of $\vb{K}$ that span the entire Brillouin zone. Consequently, we can extend our result in Eq.~\eqref{Eq:final:QO:state} to the many-body regime as
\begin{equation}
	\begin{aligned}
	\ket{\Phi_v(t)}
		&\approx \hat{\mathcal{D}}
		\bigg(
			N_z\int \dd \vb{K}\boldsymbol{\chi}_v(\vb{K},t,t_0)
		\bigg)
		\bigotimes_{q=1}^{q_c}\ket{0_q},
	\\
		&= \hat{\mathcal{D}}
			\big(
				\bar{\boldsymbol{\chi}}_v(t,t_0)
			\big)
			\bigotimes_{q=1}^{q_c}\ket{0_q},
	\end{aligned}
\end{equation}
where $N_z$ denotes the number of Brillouin zones excited by the laser field. In our numerical analysis, we consider $N_z$ to be around $10^6 - 10^7$. However, it is crucial to note that this parameter's value significantly relies on experimental conditions such as the laser beam width, the alignment of the crystal lattice relative to the field's polarization, and the sample's width in cases where the HHG process occurs during transmission.

Additionally, and as shown in Fig.~\ref{Fig:HHG:semiclassic:spec}, dephasing effects exert a pivotal influence on the final shape of the HHG spectra. We anticipate that these effects will similarly impact the generated light state. In this context, we incorporate dephasing effects phenomenologically into the computed displacements $\chi^{(q)}_v(\vb{K},t,t_0)$ while calculating the matrix elements $M^{(\text{ter})}_{\mathsf{i,j}}(\vb{K},t)$ and $M^{(\text{tra})}_{\mathsf{i,j}}(\vb{K},t)$ in Eq.~\eqref{Eq:displacements}~\footnote{This approach is somewhat equivalent to describe the crystal with a non-Hermitian Hamiltonian. In fact, this approach was studied in the context of semiclassical high-harmonic generation processes in Ref.~\cite{wang_quantum_2021}, yielding analogous outcomes to the semiclassical description provided in this text.}. This is due to their dependence on the solution of the SBEs as detailed in the previous subsection (see Appendix~\ref{Appendix:A:II} for the detailed expressions and a discussion about this matter).

\section{RESULTS}\label{Sec:Results}

\subsection{Conditioning to HHG}
In current experimental setups~\cite{lewenstein_generation_2021,lamprou_nonlinear_2023}, the creation of non-classical light states hinges on conducting anticorrelation measurements involving the part of the fundamental mode and the produced harmonics~\cite{tsatrafyllis_high-order_2017,tsatrafyllis_quantum_2019}. From a mathematical standpoint, our focus is on events where at least one photon is generated in a harmonic mode $(q\neq 1)$ while considering the correlations that arise with the fundamental mode. In mathematical terms, this involves the implementation of the projective operation~\cite{stammer_high_2022,stammer_theory_2022,stammer_quantum_2023}
\begin{equation}
	\hat{P}_{\text{cond}}
		= \mathbbm{1}-\bigotimes_{q=1}^{q_c} \dyad{0_q},
\end{equation}
which we refer to as \emph{conditioning to HHG} operation. Applying this operator to the state in Eq.~\eqref{Eq:final:QO:state} leads, up to a normalization factor, to
\begin{equation}\label{Eq:Cond:All}
	\ket{\bar{\Phi}_{v}(t)} 
		= \bigotimes^{q_c}_{q=1}
				\chiV
				- \xi_{\text{IR}}\xi_{\text{UV}} \bigotimes^{q_c}_{q=1} \ket{0_q},
\end{equation}
where we define $\xi_{\text{IR}} \equiv \langle 0_q\chiVfund$ and $\xi_{\text{UV}} \equiv \prod_{q=2}^{q_c} \langle 0_q\chiV$. The resultant state , following the application of this conditioning operation, takes the form of an entangled state encompassing all the harmonic modes excited during the HHG process, as demonstrated in Ref.~\cite{stammer_high_2022}. However, in current experimental implementations, this conditioning operation is executed by performing an anticorrelation measurement involving a portion of the outputting driving IR field and the generated harmonics~\cite{tsatrafyllis_high-order_2017,lewenstein_generation_2021,rivera-dean_strong_2022,stammer_quantum_2023}. Mathematically, this operation equates to a projection of the conditioned state in Eq.~\eqref{Eq:Cond:All} with respect to $\bigotimes_{q=2}^{q_c}\chiV$, resulting in
\begin{equation}\label{Eq:Cond:IR}
		\ket{\bar{\Phi}_{v,\text{IR}}(t)}
			= \chiVfund
				- \xi_{\text{IR}}\abs{\xi_{\text{UV}}}^2
				\ket{0_{q=1}},
\end{equation}
which is a superposition between two different coherent states. 

In the subsequent analysis, we investigate various characteristics of the generated states, including the presence of Wigner negativities, their fidelity in comparison to other states, and the degree of entanglement between the field modes. Our specific interest lies in understanding the influence of dephasing times and field strength on the computed quantities, along with the contributions from interband and intraband effects. As previously mentioned, we conduct this examination using ZnO subjected to linearly polarized light, aligned along different crystal directions, though our primary focus is on the $\Gamma-M$ direction. The laser source we consider has a Gaussian envelope, central wavelength $\lambda_L = 3.25$ $\mu$m, field strength ranging from $0.2$ to $0.6$ V/$\angstrom$, and a duration of approximately $\Delta t \sim 96$ fs (equivalent to 9 optical cycles). 

\begin{figure*}
	\centering
	\includegraphics[width=0.65\textwidth]{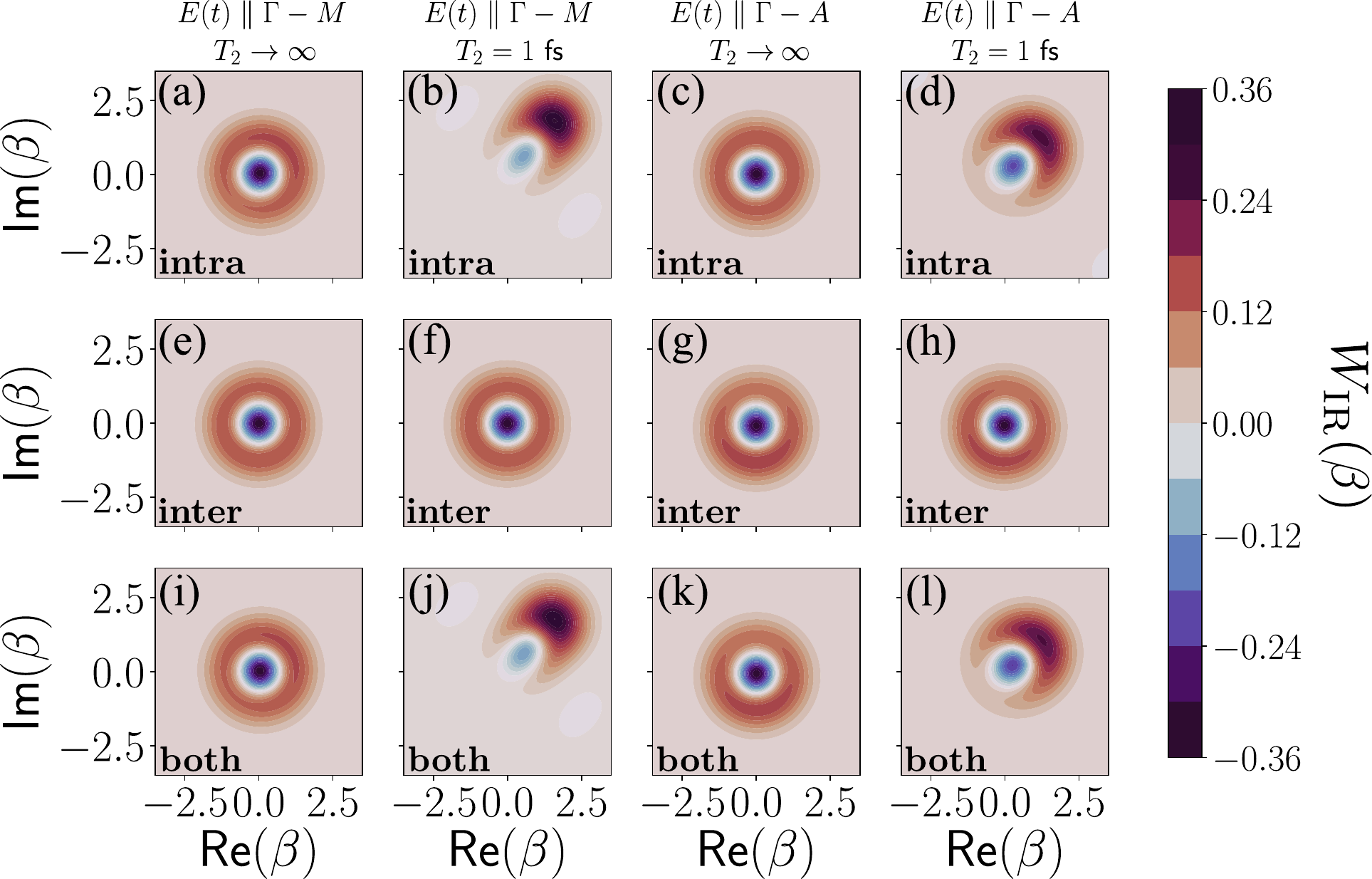}
	\caption{Wigner function representations of the state in Eq.~\eqref{Eq:Cond:IR} under various conditions. Each column corresponds to distinct values of the dephasing time, $T_2$, along with different orientations of the applied linearly polarized laser field, specifically aligned with the $\Gamma-M$ and $\Gamma-A$ directions. The first two rows focus individually on the intraband and interband contributions, while the last row displays their combined effect. The laser parameters used in this case correspond to $\lambda_L = 3.25$ $\mu$m, field strength of $0.5$ V/$\angstrom$ and a duration of $\Delta t \sim 96$ fs (equivalent to 9 optical cycles).}
	\label{Fig:Wigner:Bloch}
\end{figure*}

\subsection{Wigner function of the fundamental mode}

In Ref.~\cite{lewenstein_generation_2021}, it was demonstrated how the Wigner function of the fundamental mode could be measured in the context of HHG processes in atoms. This was achieved by employing the conditioning approach, briefly skimmed at the end of the preceding section, together with a homodyne detection scheme~\cite{smithey_measurement_1993}. In this study, our objective is to explore how the characteristics of solid-state systems might influence the resulting Wigner functions. This quasiprobability distribution, as outlined in Ref.~\cite{royer_wigner_1977}, can be computed as
\begin{equation}
	W_{\text{IR}}(\beta)
		= \tr[
				\hat{\mathcal{W}}_{q=1}(\beta)
				\dyad{\bar{\Phi}_{v,\text{IR}}(t)}
				],
\end{equation}
where we defined $\hat{\mathcal{W}}_{q}(\beta) = \hat{D}_{q}(\beta)\hat{\Pi}_{q}\hat{D}_{q}^\dagger(\beta)$ wherein $\hat{\Pi}_q$ represents the parity operator acting on mode $q$. It is noteworthy that in the case of Gaussian pure states, the resulting Wigner function maintains positivity across the entire phase space. However, this positivity does not necessarily extend to non-Gaussian states~\cite{hudson_when_1974}, as observed, for instance, with Fock states and certain coherent state superpositions, which are often used as examples of non-classical states of light. Consequently, the presence of Wigner negativities is commonly regarded as an indicator of non-classical behaviors.

In Fig.~\ref{Fig:Wigner:Bloch}, we present the obtained Wigner function under varying conditions. From left to right, in each row we explore scenarios where the laser field polarized along distinct crystal directions, using two different dephasing times: $T_2 \to \infty$~\footnote{It is worth noting that the limit $T\to\infty$ is obtained numerically by setting the dephasing time much larger than the pulse duration, such that HHG dynamics take place in a much shorter scale than dephasing effects.} and $T_2 = 1$ fs. Notably, for the diverse crystal directions, we considered distinct number of Brillouin zones. Specifically, we employed $N_z = 6.6 \times 10^6$ for the $\Gamma-M$ cases (first two columns), while opting for $N_z =1.6 \times 10^7$ for the $\Gamma-A$ direction. Bigger values of these parameters would lead to more spread and unbalanced coherent state superpositions, and therefore to Gaussian-like Wigner functions, while for smaller values we would get distributions aligning with that of displaced single-photon excitations. Moreover, these values allow us to highlight one of the main differences when considering different bands: for the same number of Brillouin zones, excitations along the $\Gamma-M$ direction lead to more spread and unbalanced coherent state superpositions compared to the $\Gamma-A$ direction. This behavior is a consequence of the bands' shape, (see Fig.~\ref{Fig:App:Params}~(b) in Appendix~\ref{App:Num:Params}) where the electron mobility along the valence band of the $\Gamma-A$ direction does not involve the same energetic exchanges as it happens for the valence band along the $\Gamma-M$ direction. Thus, less electrons, and therefore smaller $N_z$, oscillating along the $\Gamma-M$ direction are needed to reproduce the same displacements as obtained along the $\Gamma-A$ direction.

Upon comparing the columns, two conclusions can be drawn. Firstly, when $T_2$ is smaller, we observe more distinct and unbalanced coherent state superpositions with reduced Wigner negativities. This is due to the increased magnitude of $|\bar{\chi}_v^{(q=1)}(t,t_0)|$. Secondly, depending on the crystal direction the amount of displacement varies. Specifically, and as discussed earlier, for excitations along the $\Gamma-M$ direction, the displacement is bigger when using the same $N_z$. On the other hand, in the first two rows of Fig.~\ref{Fig:Wigner:Bloch}, we separately examine intraband and interband contributions, respectively, by intentionally disabling the other. The third row presents the total contribution, which is the coherent summation of both interband and intraband effects. In this scenario, a comparison across the rows highlights that the dominant contribution primarily arises from intraband terms. Furthermore, this intraband contribution tends to increase as $T_2$ decreases. 

The predominance of intraband contributions over interband contributions in the coherent state superposition formation, including the impact of dephasing times, can be elucidated as follows. As previously mentioned, intraband contributions stem from Bloch oscillations, which entail the electron's acceleration within a specific band under the influence of the field. Decreasing dephasing times signifies that the electron spends less time in the conduction band after excitation due to interactions like electron-phonon or electron-electron interactions. Consequently, the electron transitions out of the conduction band more quickly after excitation, exhibiting a non-radiative decay process. This shift in behavior makes intraband phenomena more significant compared to interband effects as $T_2$ diminishes, and also to less important contributions from interband (see Figs.~\ref{Fig:Wigner:Bloch}~(g) and (h) where the Wigner functions become slightly more homogeneous as $T_2$ decreases). Finally, this trend between interband and intraband can also be observed in the semiclassical spectra displayed in Fig.~\eqref{Fig:HHG:semiclassic:spec}, where intraband and interband contributions to the spectra become more comparable just before the perturbative region (upper bounded by the leftmost horizontal dashed line). However, it is worth noting that, while the interband component of the HHG spectra depends on the interband current, in the case of $\chi_v^{(q=1)}$ it depends on the interband polarization (see Appendix~\ref{Appendix:A:II}).

Furthermore, the shape of the conduction band, specifically its dependence on crystal momentum, significantly influences the results. This is evident in the contrasting Wigner functions obtained when considering either the $\Gamma-M$ and $\Gamma-A$ crystal directions. In such instances, the findings suggest that a flatter valence band (as illustrated in Fig.~\ref{Fig:App:Params}~(b) in Appendix~\ref{App:Num:Params}) corresponds to a smaller $|\bar{\chi}_v^{(q=1)}(t,t_0)|$. As mentioned earlier, this occurs due to the reduced energy gap between the maximum and minimum energy levels within the respective valence band, leading to a smaller energy exchange between the accelerated electron and the electromagnetic field. Thus, when $|\bar{\chi}_v^{(q=1)}(t,t_0)|$ is small enough, the first coherent state of the superposition in Eq.~\eqref{Eq:Cond:IR} can be expanded in Fock basis, such that the vacuum component cancels out with the second term of Eq.~\eqref{Eq:Cond:IR}, as in the limit $|\bar{\chi}_v^{(q=1)}(t,t_0)|\to 0$ we get $\xi_{\text{IR}}\abs{\xi_{\text{UV}}}^2 \to 1$. This leaves us with a state that is quite alike to a single-photon state, whose Wigner function has a volcano-like shape with a deep minimum at the center, similar to those shown in the central row of Fig.~\ref{Fig:Wigner:Bloch}.

\begin{figure}
	\centering
	\includegraphics[width=1\columnwidth]{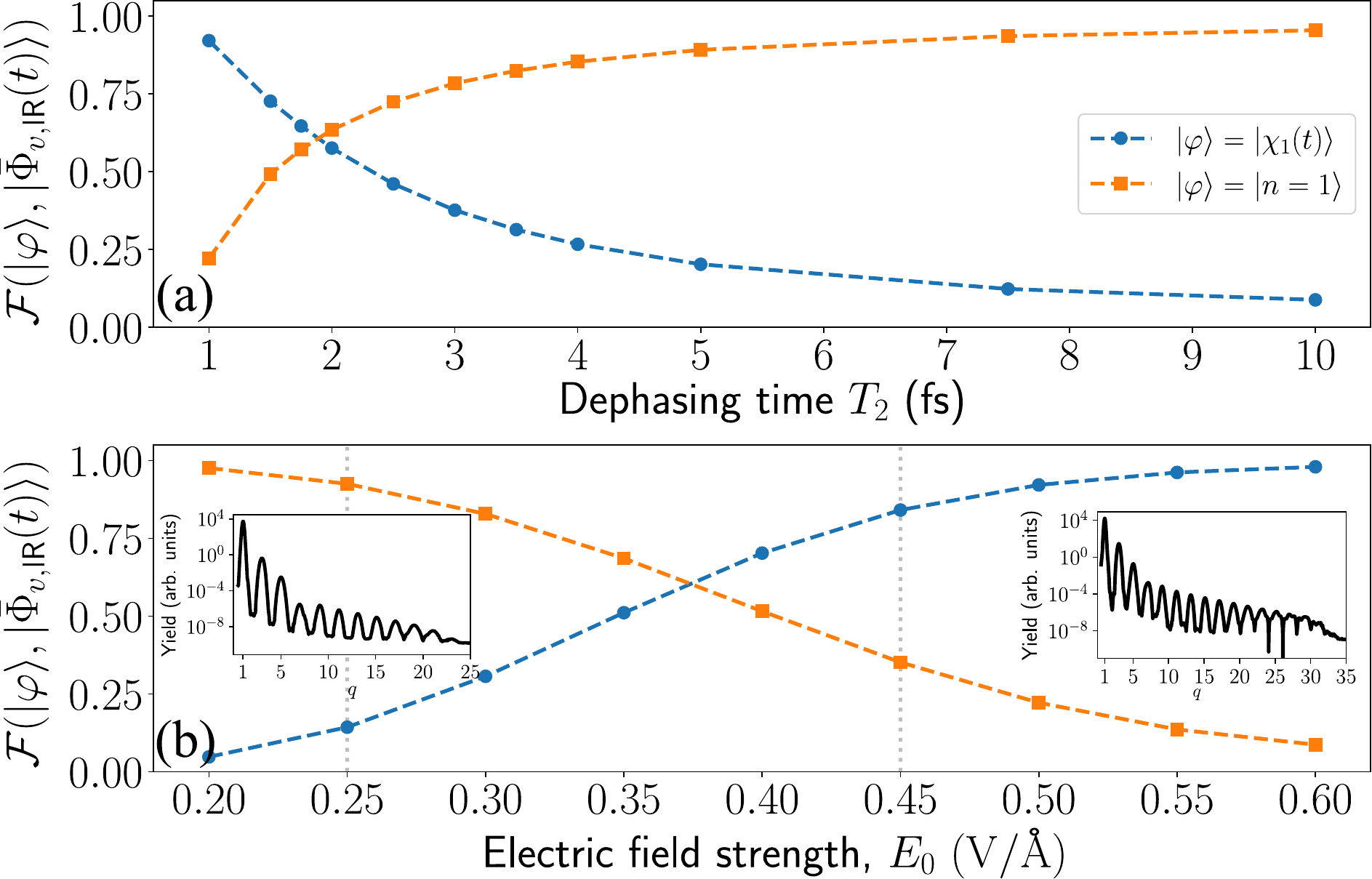}
	\caption{Fidelity of the state in Eq.~\eqref{Eq:Cond:IR} with respect to a Fock state $\ket{1_q}$ (orange curves with squared markers), and the coherent state $\chiVfund$ (blue curve with circular markers). In (a), the fidelity is plotted as a function of the dephasing time $T_2$, while in (b), it is plotted as a function of the electric field strength $E_0$. The inset plots in (b) show the semiclassical harmonic spectra, from left to right, for $E_0 = 0.25$ V$/\angstrom$ and for $E=0.45$ V$/\angstrom$. For both plots, we set the dephasing time $T_2 = 1$ fs and considered the applied laser field to be linearly polarized along the $\Gamma-M$ direction.}
	\label{Fig:Fidelities}
\end{figure}

\subsection{Fidelity of the coherent state superposition}
In previous studies~\cite{lewenstein_generation_2021,rivera-dean_strong_2022,stammer_quantum_2023}, it was demonstrated that the nature of the generated coherent state superposition through conditioning on the HHG process can vary based on the specific conditions under which the nonlinear interaction takes place. These variations range from a (displaced) Fock state, often called a \emph{kitten} state, to an unbalanced coherent state superposition, usually referred to as displaced \emph{cat} state. In the regime where $\lvert\bar{\chi}_v^{(q=1)}(t,t_0)\rvert \gg 0 $, the value $\xi_{\text{IR}}$ tends to zero, resulting in a (displaced) coherent state. 

Here, we investigate the influence of the electric field strength of the driving field and the dephasing time on the conditioned state. We achieve this by examining its fidelity in comparison to: (1) a coherent state in the form of $\chiVfund$ and (2) a Fock state $\ket{1_q}$. These two limits represent the most extreme scenarios observed in HHG within atomic systems~\cite{rivera-dean_strong_2022,stammer_quantum_2023}, as mentioned earlier. The fidelity $\mathcal{F}(\ket{\varphi},\lvert\bar{\Phi}_{v,\text{IR}}(t)\rangle)$ between $\ket{\varphi}$ and $\lvert\bar{\Phi}_{v,\text{IR}}(t)\rangle$ is given as follows
\begin{equation}
	\mathcal{F}
		\big(
			\ket{\varphi},\lvert \bar{\Phi}_{v,\text{IR}}(t)\rangle
		\big)
		\equiv \abs{\braket{\varphi}{\bar{\Phi}_{v,\text{IR}}(t)}}^2,
\end{equation}
which can be interpreted as a measure of distance between the two states~\cite{NielsenBookch9}. It is important to note that when investigating the dependence of this measure on the electric field strength, we confine our analysis to values within the range of 0.2 to 0.6 V/$\angstrom$, which are commonly employed in experimental setups~\cite{ghimire_observation_2011,schubert_sub-cycle_2014}. Values exceeding 1 V/$\angstrom$ typically surpass the damage threshold of many materials~\cite{goulielmakis_high_2022,kim_generation_2017,higley_femtosecond_2019,garratt_direct_2022}. Regarding the dephasing times, we focus on the interval 1 to 10 fs. While theoretical models~\cite{vampa_theoretical_2014,vampa_semiclassical_2015} indicate that $T_2 \sim 1$ fs is necessary to observe resolved harmonic spectra (as depicted in Fig.~\ref{Fig:HHG:semiclassic:spec}), that could potentially align with certain experimental observations in which electrons acquire high values of crystal momentum~\cite{HufnerBook}, it is generally observed that the typical dephasing times in semiconductors extend beyond 10 fs~\cite{becker_femtosecond_1988,portella_kspace_1992}.

\begin{figure*}
	\centering
	\includegraphics[width=0.75\textwidth]{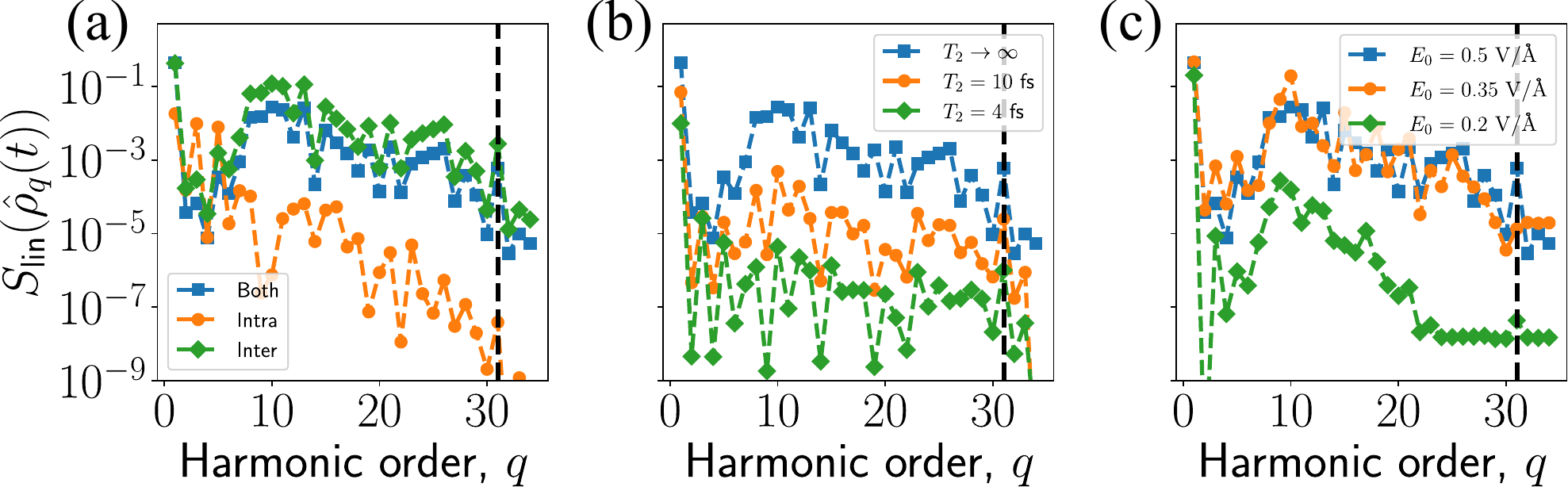}
	\caption{Linear entropy as a function of the harmonic modes. In (a), we investigate the impact of interband dynamics (green curve with diamond markers), intraband dynamics (orange curve with circular markers) and their combined effect (blue curve with squared markers) on the linear entropy. These analyses are conducted with fixed value of $E_0 = 0.5$ V/$\angstrom$ and $T_2 \to \infty$. In panel (b), we examine the influence of dephasing time using three scenarios: $T_2 \to \infty$ (blue curve with squared markers), $T_2 = 10$ fs (orange curve with circular markers) and $T_2 = 4$ fs (green curve with diamond markers), while maintaining $E_0 = 0.5$ V/$\angstrom$. Finally, in panel (c), we explore the effect of electric field strength with $T_2 \to  \infty$. We consider $E_0 = 0.5$ V/$\angstrom$ (blue curve with squared markers), $E_0 = 0.35$ V/$\angstrom$ (orange curve with circular markers) and $E_0 = 0.2$ V/$\angstrom$ (green curve with diamond markers). All cases involve a linearly polarized field along the $\Gamma-M$ direction.}
	\label{Fig:Entropy:modes}
\end{figure*}

In Fig.~\ref{Fig:Fidelities} we present the obtained results for a linearly polarized driving field aligned with the crystal direction $\Gamma-M$. Specifically, Fig.~\ref{Fig:Fidelities}~(a) illustrates the dependence of the fidelity with respect to $T_2$ (while keeping $E_0 = 0.5$ V/$\angstrom$ constant), and Fig.~\ref{Fig:Fidelities}~(b) shows the dependence on $E_0$ (with $T_2 = 1$ fs held constant).
Notably, as $T_2$ increases or $E_0$ decreases, the fidelity to the Fock state $\ket{1_q}$ rises, while the fidelity to the coherent state $\chiVfund$ decreases. Although dephasing times are largely beyond experimental control as they are due to material properties, the electric field strengh offers a means to influence the fidelity of the generated state. 

Moreover, within the considered ranges --consistent with experimental implementations-- the generated state's fidelity is highly tunable. For instance, the fidelity to the Fock (coherent state) changes from approximately $\mathcal{F} \approx 0.98$ ($\approx0.05$) at $E_0 = 0.2$ V/$\angstrom$, to around $\mathcal{F} \approx 0.09$ ($\approx0.98$) at $E_0 = 0.6$ V/$\angstrom$, since the smaller (bigger) $\lvert\bar{\chi}^{(q=1}(t,t_0)\rvert$ is, the more similar the genated state would be to a Fock (coherent) state. It is crucial to note that our analysis has been based on $N_z= 6\times 10^6$, implying that factors such as the beam width in the interaction region and the sample's width (in the case of HHG during transmission) also play an instrumental role in controlling the generated states fidelity. 

\subsection{Entanglement between the field modes}
As mentioned in Section~\ref{Sec:QO:des}, the state presented in Eq.~\eqref{Eq:Cond:All} takes the form of an entangled state among all the field modes excited during the HHG process. This was demonstrated in a previous study~\cite{stammer_high_2022}, where the degree of entanglement between one mode $q$ and all the others ($\forall q' \neq q$) was investigated as a function of $\lvert \bar{\chi}_v^{(q=1)}(t,t_0)\rvert$. To accomplish this, the linear entropy~\cite{agarwal_quantitative_2005,berrada_beam_2013} was employed as entanglement measure
\begin{equation}
	S_{\text{lin}}(\hat{\rho}_q) 
		\equiv 1- \tr(\hat{\rho}_q^2), 
\end{equation}
which is obtained through a first-order expansion of the von Neumann entropy~\cite{berrada_beam_2013},
where $\hat{\rho_q} = \tr_{q'\neq q}(\dyad{\bar{\Phi}_v(t)})/\tr(\dyad{\bar{\Phi}_v(t)})$. It is worth noting that, unlike the von Neumann entropy, the linear entropy is bounded as $0 \leq S_{\text{lin}}(\hat{\rho}_q) \leq 0.5$, the lower bound obtained for separable states ($\hat{\rho}_q$ being a pure state) while the upper bound for maximally entangled states ($\hat{\rho}_q$ being a maximally mixed state). This method is particularly suitable for computations involving continuous variable pure states, such as coherent states.

Here, we investigate how this entanglement measure changes for different harmonic modes under various conditions. This is illustrated in Fig.~\ref{Fig:Entropy:modes}, where similar features akin to a typical harmonic spectrum are evident. A frequency plateau extends until a sharp cutoff, which, when the field intensity is sufficiently high, aligns with the maximum energy bandgap (indicated by the dashed vertical black curve).

In particular, Fig.~\ref{Fig:Entropy:modes}~(a) considers a field with amplitude $E_0 = 0.5$ $\mathrm{V}/\angstrom$, while setting $T\to \infty$, to examine the contributions of interband (blue curve with squared markers) and intraband (orange curve with circular markers) effects to the linear entropy, as well as their combined contribution (green line with diamond markers). In contrast to the Wigner function distribution, in this case interband effects dominate. Notably, for $q=1$, the amount of entanglement reaches a maximum value of $S_{\text{lin}}(\hat{\rho}_{q=1}) = 0.44$. This dominance aligns with expectations, as revealed in Fig.~\ref{Fig:HHG:semiclassic:spec}, where within the perturbative part the spectrum, interband effects play a more pivotal role than intraband effects, influencing both the shape of the harmonic plateau and the cutoff frequency. 

Conversely, as the dephasing time decreases, the harmonic yield for constant field ($E_0=0.5$ V/$\angstrom$) diminishes, resulting in a reduction in the total entanglement as observed in Fig.~\ref{Fig:Entropy:modes}~(b). In Fig.~\ref{Fig:Entropy:modes}~(c), we investigate the variation in the degree of entanglement as a function of the harmonic mode when considering different field strengths and assuming the limit $T_2 \to \infty$. It can be observed that, for lower field strengths, the cutoff frequency at which the linear entropy reaches zero decreases. This behavior is analogous to the pattern seen in the harmonic spectrum (as depicted in the insets of Fig.~\ref{Fig:Fidelities} and further discussed in Ref.~\cite{ghimire_observation_2011}). On the other hand, we find that the amount of entanglement within the plateau region diminishes as the field strength decreases.

\begin{figure}
	\centering
	\includegraphics[width=1\columnwidth]{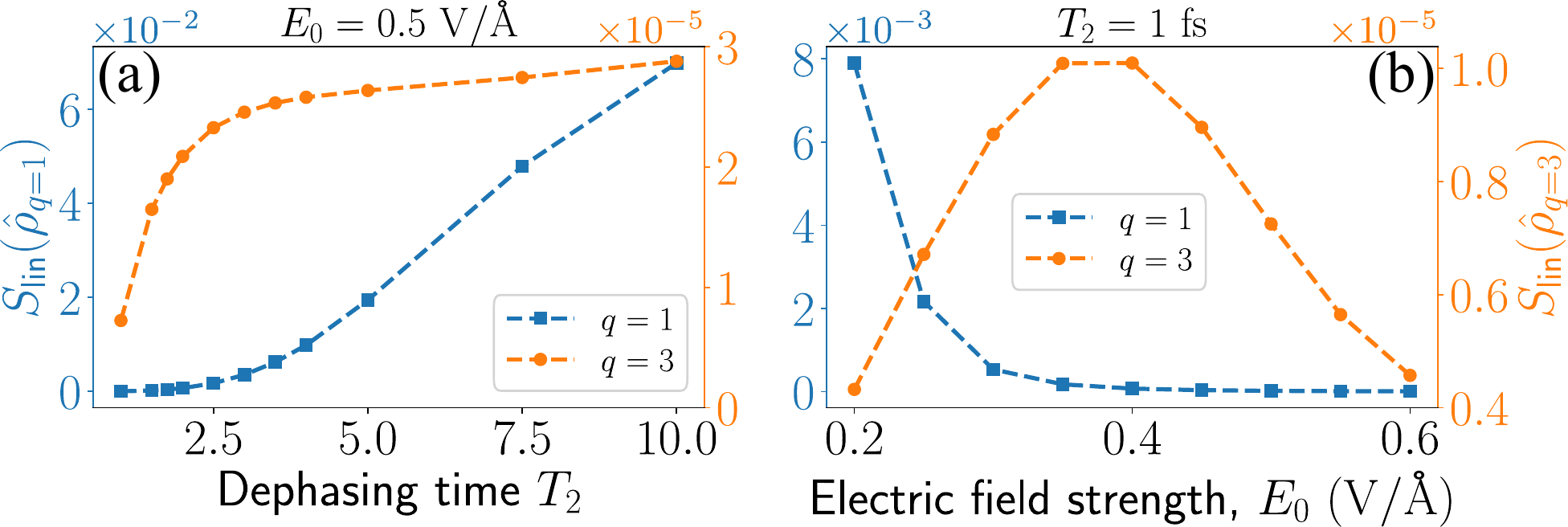}
	\caption{Dependence of the linear entropy on (a) dephasing time and (b) electric field strength for the first (blue curve with squared markers) and third (orange curve with circular markers) harmonic modes. In panel (a) we maintain $E_0 = 0.5$ V/$\angstrom$ while in panel (b) we fixed $T_2 = 1$ fs constant. In both scenarios, the incident light is linearly polarized light along the $\Gamma-M$ direction.}
	\label{Fig:Entropy:params}
\end{figure}

The trends observed in Fig.~\ref{Fig:Entropy:modes} are further highlighted in Fig.~\ref{Fig:Entropy:params} with some remarkable differences. In Fig.~\ref{Fig:Entropy:params} the linear entropy is shown as a function of the dephasing time for the same field strength value $E_0 = 0.5$ V/$\angstrom$. Notably, both the first (blue curve with squared markers) and the third (orange curve with circular markers) harmonics exhibit an increase in the linear entropy with higher $T_2$. However, it is noteworthy that the rate of change in entropy varies between the first and third harmonic modes. On the other hand, in Fig.~\ref{Fig:Entropy:params}~(b) we show the dependence of the linear entropy as a function of the field strength for a constant dephasing time $T_2 = 1$ fs for the aforementioned harmonic modes. These two harmonic follow different trends, at very distinct orders of magnitude, on its behavior for increasing field strengths. While for $q=1$ the degree of entanglement decreases with increasing field strengths, for $q=3$ this quantity reaches a maximum around $E_0 \approx 0.38$ V/$\angstrom$. Notably, we observe that dephasing time influences on the dependence of degree of entanglement for harmonic modes, as in Fig.~\ref{Fig:Entropy:modes}~(c) it is found that decreasing field strengths lead to reduced values of the linear entropy for the considered harmonic modes.

Although in this instance, entanglement features among distinct modes emerge due to the conditioning on HHG operation, recent findings~\cite{stammer_squeezing_2023} in atomic systems demonstrate that beyond the weak-depletion approximation of the ground state, squeezing and entanglement features can manifest in harmonic modes without conditioning. Hence, an avenue for further exploration could entail investigating a similar scenario in solids. Specifically, Ref.~\cite{r_temporal_2023} showcased that the utilization of fields with a $\text{sinc}^2$ envelope necessitates considerations beyond the weak-depletion approximation of the valence band. Consequently, one could think about the use of this kind of pulses to observe similar features in solid-state systems.

 \section{CONCLUSIONS AND OUTLOOK}\label{Sec:Conclusions}
In this work we have investigated high-harmonic generation processes in semiconductor materials under a quantum optical perspective. Our study focused on the interaction of a strong infrared laser field, linearly polarized along different crystal orientations of a ZnO material, which has been extensively examined both experimental (e.g.~Ref.~\cite{ghimire_observation_2011}) and theoretically (e.g.~Ref.~\cite{vampa_theoretical_2014}). Similar to atomic systems~\cite{lewenstein_generation_2021,rivera-dean_strong_2022,stammer_high_2022,stammer_quantum_2023}, our findings show that conditioning measurements on the high-harmonic generation process, enable for the creation of entangled coherent state superpositions. Thus, these results suggest that the experiments conducted in Refs.~\cite{lewenstein_generation_2021,rivera-dean_strong_2022,lamprou_nonlinear_2023} could potentially be conducted at smaller length scales, considering that typical semiconductor samples, utilized in much of the current technology, are very small~\cite{ghimire_observation_2011,schubert_sub-cycle_2014}.

We observed that the generated states are strongly influenced by the electron interband and intraband dynamics, excitation conditions including the polarization direction and field strength, and inherent solid-state properties such as the dephasing time. Specifically, the analysis of non-classical properties on the final IR state, assessed through the presence of Wigner function negativities, reveals that intraband dynamics play a predominant role, and get reduced for increasing dephasing times and for decreasing field strengths. Conversely, the evaluation of entanglement properties, characterized by the linear entropy, indicates that interband dynamics have a more pronounced impact in this case. Enhanced values of this entanglement measure are observed with increasing dephasing times while the behavior with respect to the field strength depends on the targeted harmonic mode and the considered dephasing time. Thus, these two non-exclusive indications of non-classical behaviors appear in two opposite regimes. These results suggest that, in experimental implementations, obtaining Wigner function similar to those presented in this text would be more accessible than finding non-negligible values of entanglement between the different field modes. This is because the inclusion of dephasing times do not eliminate the quantum superposition between the states, as they can also be countered with other controllable experimental parameters such as the width of the solid-state sample and the laser field strength.

While we have observed that dephasing effects strongly influence the non-classical properties of the state, it is important to emphasize that the role of dephasing times and the values typically considered in theoretical analyses remain subjects of ongoing debate within the research community. In contrast to certain theoretical findings, some experimental results suggest dephasing times beyond 10 fs, while other recent theoretical findings propose that long trajectories in real space of solids can lead to destructive interference, resulting in clear harmonic spectra even in the regime of $T_2 \sim 10$ fs~\cite{brown_real-space_2022}. Additionally and beyond dephasing, propagation effects and other macroscopic aspects~\cite{floss_ab_2018,abadie_spatiotemporal_2018} also contribute to shaping the features of the resulting harmonic spectra. As such, we anticipate that these various factors play crucial roles in determining the ultimate properties of the generated non-classical states, extending beyond the scope of the considerations discussed in this work.

Finally, it is worth noting that our work involves certain approximations. One such approximation pertains to the consideration of a discrete set of modes and the incorporation of a weighting function to account for the pulse envelope when multiplying the electric field operator. While this approach has proven its validity in explaining different experimental observations, as seen in references such as~\cite{lewenstein_generation_2021,rivera-dean_strong_2022,stammer_quantum_2023,lamprou_nonlinear_2023}, we have observed that in solid-state systems, it may potentially impact the entanglement features between light and matter. Therefore, a natural progression would involve exploring a more precise description that encompasses a continuous set of electromagnetic field modes. Furthermore, our analysis has primarily focused on the effects of intraband and interband dynamics, while neglecting the influence of other factors like those associated with mixed currents. Although these aspects remain open questions from a semiclassical perspective, a promising avenue for further research lies in investigating how these effects, along with contributions from Berry connection terms, shape the quantum optical state.

\section*{ACKNOWLEDGMENTS}
ICFO group acknowledges support from: ERC AdG NOQIA; Ministerio de Ciencia y Innovation Agencia Estatal de Investigaciones (PGC2018--097027--B--I00 / 10.13039 / 501100011033, CEX2019--000910--S / 10.13039 / 501100011033, Plan National FIDEUA PID 2019--106901GB--I00, FPI, QUANTERA MAQS PCI 2019--111828--2, QUANTERA DYNAMITE PCI 2022--132919,  Proyectos de I+D+I “Retos Colaboración” QUSPIN RTC 2019--007196--7); MICIIN with funding from European Union Next Generation EU (PRTR--C17.I1) and by Generalitat de Catalunya;  Fundació Cellex; Fundació Mir-Puig; Generalitat de Catalunya (European Social Fund FEDER and CERCA program, AGAUR Grant No. 2021 SGR 01452, QuantumCAT \ U16--011424, co-funded by ERDF Operational Program of Catalonia 2014-2020); Barcelona Supercomputing Center MareNostrum (FI--2022--1--0042); EU Horizon 2020 FET--OPEN OPTOlogic (Grant No 899794); EU Horizon Europe Program (Grant Agreement 101080086 — NeQST), National Science Centre, Poland (Symfonia Grant No. 2016/20/W/ST4/00314); ICFO Internal “QuantumGaudi” project; European Union’s Horizon 2020 research and innovation program under the Marie-Skłodowska-Curie grant agreement No 101029393 (STREDCH) and No 847648  (“La Caixa” Junior Leaders fellowships ID100010434 : LCF / BQ / PI19 / 11690013, LCF / BQ / PI20 / 11760031, LCF / BQ / PR20 / 11770012, LCF / BQ / PR21 / 11840013). Views and opinions expressed in this work are, however, those of the author(s) only and do not necessarily reflect those of the European Union, European Climate, Infrastructure and Environment Executive Agency (CINEA), nor any other granting authority.  Neither the European Union nor any granting authority can be held responsible for them. 

P. Tzallas group at FORTH acknowledges LASERLABEUROPE V (H2020-EU.1.4.1.2 grant no.871124), FORTH Synergy Grant AgiIDA (grand no. 00133), the H2020 framework program for research and innovation under the NEP-Europe-Pilot project (no. 101007417). ELI-ALPS is supported by the European Union and co-financed by the European Regional Development Fund (GINOP Grant No. 2.3.6-15-2015-00001).

J.R-D. acknowledges funding from the Secretaria d'Universitats i Recerca del Departament d'Empresa i Coneixement de la Generalitat de Catalunya, the European Social Fund (L'FSE inverteix en el teu futur)--FEDER, the Government of Spain (Severo Ochoa CEX2019-000910-S and TRANQI), Fundació Cellex, Fundació Mir-Puig, Generalitat de Catalunya (CERCA program) and the ERC AdG CERQUTE. 

P.S. acknowledges funding from the European Union’s Horizon 2020 research and innovation programme under the Marie Skłodowska-Curie grant agreement No 847517. 

A.S.M. acknowledges funding support from the European Union’s Horizon 2020 research and innovation programme under the Marie Sk{\l}odowska-Curie grant agreement, SSFI No.\ 887153.

E.P. acknowledges the Royal Society for University Research Fellowship funding under URF$\setminus$R1$\setminus$211390.

M.F.C. acknowledges financial support from the Guangdong Province Science and Technology Major Project (Future functional materials under extreme conditions - 2021B0301030005) and the Guangdong Natural Science Foundation (General Program project No. 2023A1515010871).

\bibliography{References.bib}{}
\clearpage
\onecolumngrid
\appendix
\renewcommand\thefigure{\thesection.\arabic{figure}}
\begin{center}
	{\large \bf \textsc{Appendix}}
\end{center}

\section{About ZnO parameters and the numerical calculations}\label{App:Num:Params}
\setcounter{figure}{0}  
For the numerical simulations we have done in this text, we considered as solid-state system crystalline ZnO, which is a centrosymmetric material with a cubic lattice structure. In order to model it, we considered a two-band, tight-binding model with energy-dispersion relations for valence and conduction band defined in Eqs.~\eqref{Eq:val:dis} and \eqref{Eq:cond:dis}. The expansion coefficients $\alpha^j_{m,i}$ have been extracted from Ref.~\cite{vampa_semiclassical_2015,vampa_high-harmonic_2015}, and can be calculated using the nonlocal empirical pseudopotential method (NL-EPM), see e.g. Ref.~\cite{goano_electronic_2007}. For the sake of completeness, we show these values in Table.~\ref{Table:Params}, as well as the corresponding lattice constants along the different directions, which we refer here as $\Gamma-M \to x$, $\Gamma-K \to y$ and $\Gamma-A\to z$. In Fig.~\ref{Fig:App:Params} we plot these energy band dispersion relations for the different solid directions. Here, we restricted to electrons that are around $\Gamma$ point, but we allow for freedom in the value of crystal momentum that is being plotted, $k_i$. In our numerical calculations, we also considered values of the crystal momentum close to the $\Gamma$ point for those solid directions along which the laser field is not parallel. Increasing values of $\vb{k}$ (within the corresponding Brillouin zone) along these directions leads to smaller contributions in the HHG spectrum.

\begin{table}[h!]
	\begin{tabular}{|l|l|lll}
		\hline
		\multicolumn{1}{|c|}{\textbf{Valence band}} & \multicolumn{1}{c|}{\textbf{Conduction band}} & \multicolumn{3}{c|}{\textbf{Lattice constants} (in a.u.)}                                                                         \\ \hline
		$\alpha^0_{v,x} = -0.0928$         & $\alpha^0_{c,x} = 0.0898$            & \multicolumn{1}{l|}{$a_x = 5.32$}      & \multicolumn{1}{l|}{$a_y = 6.14$}      & \multicolumn{1}{l|}{$a_z = 9.83$}      \\ \cline{3-5} 
		$\alpha^1_{v,x} = 0.0705$          & $\alpha^1_{c,x} = -0.0814$           & \multicolumn{3}{c|}{\textbf{Kane parameters} (in a.u.)}                                                                           \\ \cline{3-5} 
		$\alpha^2_{v,x} = 0.0200$          & $\alpha^2_{c,x} = -0.0024$           & \multicolumn{1}{l|}{$E_{p,x} = 0.479$} & \multicolumn{1}{l|}{$E_{p,y} = 0.355$} & \multicolumn{1}{l|}{$E_{p,z} = 0.355$} \\ \cline{3-5} 
		$\alpha^3_{v,x} = -0.0012$         & $\alpha^3_{c,x} = -0.0048$           & \multicolumn{3}{l}{\multirow{7}{*}{}}                                                                                    \\
		$\alpha^4_{v,x} = 0.0029$          & $\alpha^4_{c,x} = -0.0003$           & \multicolumn{3}{l}{}                                                                                                     \\
		$\alpha^5_{v,x} = 0.0006$          & $\alpha^5_{c,x} = -0.0009$           & \multicolumn{3}{l}{}                                                                                                     \\ \cline{1-2}
		$\alpha^0_{v,y} = -0.0307$         & $\alpha^0_{c,y} = 0.1147$            & \multicolumn{3}{l}{}                                                                                                     \\
		$\alpha^1_{v,y} = 0.0307$          & $\alpha^1_{c,y} = -0.1147$           & \multicolumn{3}{l}{}                                                                                                     \\ \cline{1-2}
		$\alpha^0_{v,z} = -0.0059$         & $\alpha^0_{c,z} = 0.0435$            & \multicolumn{3}{l}{}                                                                                                     \\
		$\alpha^1_{v,z} = 0.0059$          & $\alpha^1_{c,z} = -0.0435$           & \multicolumn{3}{l}{}                                                                                                     \\ \cline{1-2}
	\end{tabular}
	\caption{Solid parameters used to model crystalline ZnO. The data we used has been extracted from Refs.~\cite{vampa_theoretical_2014,vampa_semiclassical_2015,vampa_high-harmonic_2015}.}
	\label{Table:Params}
\end{table}

On the other hand, and unlike the numerical analysis done in Refs.~\cite{vampa_theoretical_2014,vampa_semiclassical_2015}, in our description we do account for the dependence with the crystal momentum of the dipole matrix elements between valence and conduction bands. In terms of the $\vb{k}\cdot \vb{p}$ approximation~\cite{HaugBookch5}, these can be written as
\begin{equation}
	d^{(i)}_{vc}(\vb{k}) = \sqrt{\dfrac{E_{p,i}}{[2(E_c(\vb{k})-E_v(\vb{k}))^2]}},
\end{equation}
and which are shown in Fig.~\ref{Fig:App:Params}~(a) as a function of the crystal momentum. Here, we also considered values around the $\Gamma$ point for the solid directions different to those that are being actually plotted.
\begin{figure*}[h!]
	\centering
	\includegraphics[width = 0.75\textwidth]{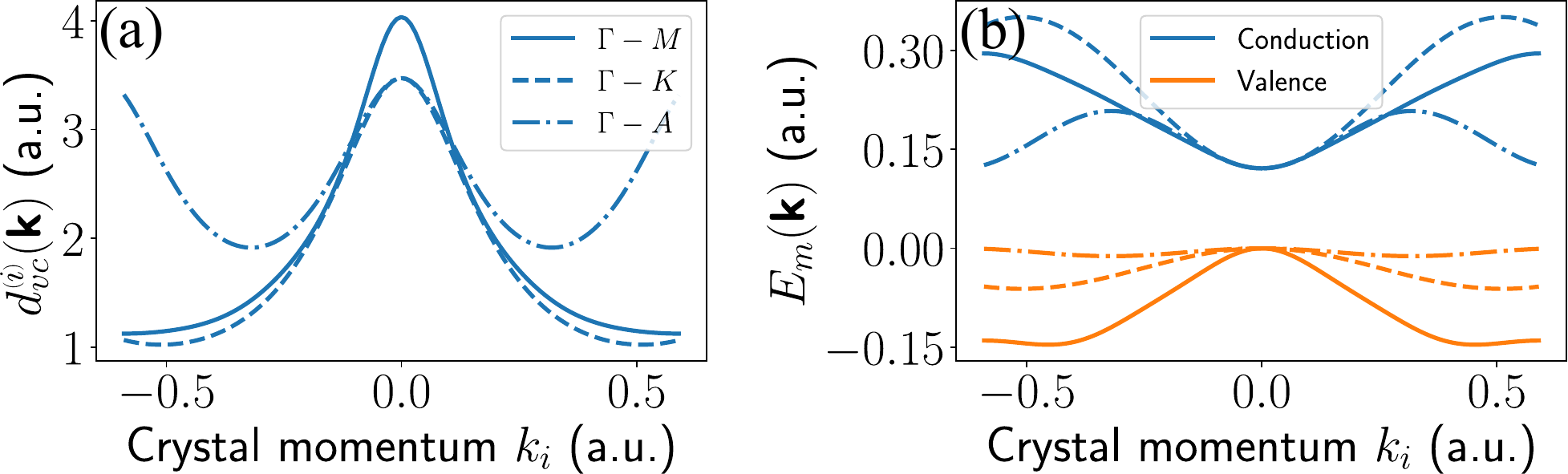}
	\caption{In (a), the transition dipole matrix element between valence and conduction band is plotted for the different solid directions. In (b), the same is done for the dispersion relation. Each curve corresponds to a different solid direction. These plots have been done for the $\Gamma$ point, which means that when plotting these functions for each solid direction, we have set the other values of crystal momentum to zero.}
	\label{Fig:App:Params}
\end{figure*}

The numerical simulations presented in this text have been done in \texttt{Mathematica} and using a MacBook Air M1 (2020) with 8 GPU cores. Once the solid and  laser field parameters are fixed, each simulation using all the cores in parallel takes around 5 minutes to finalize.

\section{Step-by-step derivation of the quantum optical description under the single-active electron approximation}
\setcounter{figure}{0}  
In this appendix, we present a comprehensive derivation of the equations outlined in the main text about the quantum optical description of the laser-solid interaction. We begin the derivation from Eq.~\eqref{Eq:H:QO:I} and proceed step by step. Initially, in Sec.~\ref{Appendix:A:I} we introduce certain transformations to facilitate subsequent calculations, and we further make specific approximations. Following this, in Sec.~\ref{Appendix:A:II} we advance to solve the resultant Schrödinger equation.

\subsection{Transformations and approximations}\label{Appendix:A:I}
Starting from the definition of the semiclassical Hamiltonian given in Eq.~\eqref{Eq:H:sc}, we can rewrite Eq.~\eqref{Eq:H:QO:I} as follows
\begin{equation}
	i\hbar \pdv{\ket{\bar{\psi}(t)}}{t}
		= \Big[
				\hat{H}_{\text{cr}}
				+ \mathsf{e}\hat{r}_{i,\text{tra}}
					\big(
						E_{\text{cl}}(t) + \hat{E}(t)
					\big)
				+  \mathsf{e}\hat{r}_{i,\text{ter}}
				\big(
				E_{\text{cl}}(t) + \hat{E}(t)
				\big)
			\Big]
				\ket{\bar{\Psi}(t)},
\end{equation}
and introducing the transformation 
\begin{equation}
	\ket{\bar{\Psi}(t)}
		= e^{i\mathsf{e}(A_{\text{cl}}(t) + \hat{A}(t))\hat{r}_{i,\text{tra}}/\hbar}
		\ket{\bar{\Psi}'(t)},
\end{equation}
where $A_{\text{cl}}(t) = \text{tr}(\hat{A}(t) \dyad{\Psi(t_0)})$ with $\hat{A}(t)$ the vector potential operator defined as $\hat{E}(t) = -\partial\hat{A}(t)/\partial t$, we get 
\begin{equation}
	\begin{aligned}
	&-\mathsf{e}\pdv{}{t}
		\big(
			A_{\text{cl}}(t)+\hat{A}(t)
		\big)
		\hat{r}_{i,\text{tra}}
		  e^{i\mathsf{e}(A_{\text{cl}}(t) + \hat{A}(t))\hat{r}_{i,\text{tra}}/\hbar}
		  \ket{\Psi'(t)}
	+ i\hbar 
		e^{i\mathsf{e}(A_{\text{cl}}(t) + \hat{A}(t))\hat{r}_{i,\text{tra}}/\hbar}
		\pdv{\ket{\bar{\Psi}'(t)}}{t}
	\\&\hspace{2cm}
		= \Big[
				\hat{H}_{\text{cr}}
				+ \mathsf{e}\hat{r}_{i,\text{tra}}
				\big(
					E_{\text{cl}}(t) + \hat{E}(t)
				\big)
				+  \mathsf{e}\hat{r}_{i,\text{ter}}
				\big(
					E_{\text{cl}}(t) + \hat{E}(t)
				\big)
			\Big]
		e^{i\mathsf{e}(A_{\text{cl}}(t) + \hat{A}(t))\hat{r}_{i,\text{tra}}/\hbar}
		\ket{\bar{\Psi}(t)},
	\end{aligned}
\end{equation}
which after some reordering, leads to
\begin{equation}\label{Eq:App:Sch:QO:II}
	i\hbar \pdv{\ket{\bar{\Psi}'(t)}}{t}
		= e^{-i\mathsf{e}(A_{\text{cl}}(t) + \hat{A}(t))\hat{r}_{i,\text{tra}}/\hbar}
			\Big[
					\hat{H}_{\text{cr}}
				+  \mathsf{e}\hat{r}_{i,\text{ter}}
				\big(
				E_{\text{cl}}(t) + \hat{E}(t)
				\big)
			\Big]
			e^{i\mathsf{e}(A_{\text{cl}}(t) + \hat{A}(t))\hat{r}_{i,\text{tra}}/\hbar}
			\ket{\bar{\Psi}'(t)}.
\end{equation}

The coupling factor $\text{g}(\omega_q)$ in the vector potential's definition is relatively small, rendering quantum optical fluctuations a perturbative element (at the level of single electrons). Consequently, if we expand $e^{i\mathsf{e}\hat{A}(t)\hat{r}_{i,\text{tra}}/\hbar}$ as a polynomial series, retaining terms up to first order in $\text{g}(\omega_q)$, i.e., $e^{i\mathsf{e}\hat{A}(t)\hat{r}_{i,\text{tra}}/\hbar} \approx \mathbbm{1} + i\mathsf{e}\hbar^{-1} \hat{A}(t)\hat{r}_{i,\text{tra}}$, we obtain, for each term in Eq.~\eqref{Eq:App:Sch:QO:II}, the following expressions
\begin{align}
	&e^{-i\mathsf{e} \hat{A}(t) \hat{r}_{i,\text{tra}}/\hbar}
		\hat{H}_{\text{cr}}
	e^{i \mathsf{e}\hat{A}(t) \hat{r}_{i,\text{tra}}/\hbar}
	\approx 
	\hat{H}_{\text{cr}}
	- i\mathsf{e}\hbar^{-1} \hat{A}(t) \hat{r}_{i,\text{tra}}
		\hat{H}_{\text{cr}}
	+ i\mathsf{e}\hbar^{-1} \hat{A}(t) \hat{H}_{\text{cr}} \hat{r}_{i,\text{tra}}\nonumber
	\\&\hspace{4.77cm}
	= \hat{H}_{\text{cr}}
		-i\mathsf{e}\hbar^{-1}\hat{A}(t) \comm{\hat{r}_{i,\text{tra}}}{\hat{H}_{\text{cr}}},
	\label{Eq:App:Comm:Hcr}
	\\ 
	&
	e^{-i\mathsf{e} \hat{A}(t) \hat{r}_{i,\text{tra}}/\hbar}
	\hat{r}_{i,\text{ter}}
	e^{i \mathsf{e}\hat{A}(t) \hat{r}_{i,\text{tra}}/\hbar}
	\approx 
	\hat{r}_{i,\text{ter}}
	- i\mathsf{e}\hbar^{-1} \hat{A}(t) \hat{r}_{i,\text{tra}}
	\hat{r}_{i,\text{ter}}
	+ i\mathsf{e}\hbar^{-1} \hat{A}(t)\hat{r}_{i,\text{ter}} \hat{r}_{i,\text{tra}}\nonumber
	\\&\hspace{4.93cm}
	= \hat{r}_{i,\text{ter}}
	-i\mathsf{e}\hbar^{-1}\hat{A}(t) \comm{\hat{r}_{i,\text{tra}}}{\hat{r}_{i,\text{ter}}},
	\label{Eq:App:Comm:rter}
	\\&
	e^{-i\mathsf{e} \hat{A}(t) \hat{r}_{i,\text{tra}}/\hbar}
	\hat{E}(t)
	e^{i \mathsf{e}\hat{A}(t) \hat{r}_{i,\text{tra}}/\hbar}
	\approx 
	\hat{E}(t)
	- i\mathsf{e}\hbar^{-1}  \hat{r}_{i,\text{tra}}\hat{A}(t)
	\hat{E}(t)
	+ i\mathsf{e}\hbar^{-1} \hat{r}_{i,\text{tra}}\hat{E}(t)\hat{A}(t)\nonumber
	\\&\hspace{4.9cm}
	\approx \hat{E}(t),
	\label{Eq:App:Comm:E}
\end{align}
where, in Eq.~\eqref{Eq:App:Comm:E}, we have additionally neglected terms of the form $\hat{E}(t)\hat{A}(t)$, which are second order in terms of $\text{g}(\omega_q)$. It's worth noting that by making this simplification, we exclude second-order terms involving creation and annihilation operators. These terms could potentially introduce entanglement between the field modes and even lead to squeezing effects~\cite{stammer_iwo_2023}.

Moving on to the other two equations, we observe that Eq.\eqref{Eq:App:Comm:Hcr} introduces the influence of the intraband current, as defined in Eq.\eqref{Eq:intra:curr}, while Eq.~\eqref{Eq:App:Comm:rter} captures the impact of the mixture current. As highlighted in the main text, the primary objective of this work is to examine the backaction of interband and intraband currents on the quantum optical state. As such, we omit the consideration of the effects of the mixture current, which still remain uncertain within the scope of semiclassical theories.

Thus, under the previous approximations, the Schrödinger equation we aim to solve is
\begin{equation}
	i\hbar \pdv{\ket{\bar{\Psi}'(t)}}{t}
		= 
			\Big[
				\hat{H}_{\text{cr}}(t)
				+ \mathsf{e}\hat{r}_{i,\text{ter}}(t)
					\big(
						E_{\text{cl}}(t)
						+ \hat{E}(t) 
					\big)
				- i\mathsf{e}\hbar^{-1}\hat{v}_{i,\text{tra}}(t) \hat{A}(t)
			\Big]  \ket{\bar{\Psi}'(t)},
\end{equation}
where we have defined $\hat{v}_{i,\text{tra}} \equiv [\hat{r}_{i,\text{tra}},\hat{H}_{\text{cr}}]$. Note that in this expression, $\hat{H}(t)$ and $\hat{r}_{i,\text{ter}}(t)$ are defined as in Eqs.~\eqref{Eq:Hcr:transf} and \eqref{Eq:rter:transf}, and $\hat{v}_{i,\text{tra}}(t)$ is obtained similarly as in these equations. At this point, we move to the interaction picture with respect to $\hat{\bar{H}}_{\text{sc}}(t) \equiv \hat{H}_{\text{cr}}(t)
+ \mathsf{e}\hat{r}_{i,\text{ter}}E_{\text{cl}}(t)$, that is
\begin{equation}
	\ket{\bar{\Psi}'(t)}
		= \hat{U}_{\text{sc}}(t,t_0) \tPsi,
\end{equation}
where $\hat{U}_{\text{sc}}(t,t_0) \equiv \hat{\mathcal{T}}\exp[-i\int^t_{t_0} \dd \tau \hat{\bar{H}}_{\text{sc}}(\tau)/\hbar]$ with $\hat{\mathcal{T}}$ the time-ordering operator. It is important to note here that this unitary transformation is the one solving Eq.~\eqref{Eq:Hcr:transf}. Therefore, the obtained Schrödinger equation reads as
\begin{equation}\label{Eq:App:Sch:Q:final}
		i\hbar \pdv{\tPsi}{t}
	= 
	\Big[
		\mathsf{e}\hat{\bar{r}}_{i,\text{ter}}(t)
			\hat{E}(t) 
	- i\mathsf{e}\hbar^{-1}\hat{\bar{v}}_{i,\text{tra}} \hat{A}(t)
	\Big]  \tPsi,
\end{equation}
where we have defined $\hat{\bar{r}}_{i,\text{ter}}(t) \equiv \hat{U}^\dagger_{\text{sc}}(t)\hat{r}_{i,\text{ter}}(t)\hat{U}_{\text{sc}}(t)$ and  $\hat{\bar{v}}_{i,\text{tra}}(t)  \equiv \hat{U}^\dagger_{\text{sc}}(t)\hat{v}_{i,\text{tra}}(t)\hat{U}_{\text{sc}}(t)$.

\subsection{Solving the obtained equation}\label{Appendix:A:II}
To solve the equation presented in Eq.~\eqref{Eq:App:Sch:Q:final}, we project it onto $\ket{\vb{K},m}$, resulting in
\begin{equation}\label{Eq:App:Sch:proj}
	i\hbar \pdv{}{t}\ket{\Phi_m(\vb{K},t)}
		= \bra{\vb{K},m}
				\Big[
			\mathsf{e}\hat{\bar{r}}_{i,\text{ter}}(t)
			\hat{E}(t) 
			- i\mathsf{e}\hbar^{-1}\hat{\bar{v}}_{i,\text{tra}} \hat{A}(t)
			\Big]  \tPsi,
\end{equation}
where we have defined $\ket{\Phi_m(\vb{K},t)} = \langle\vb{K},m\tPsi$. Introducing the identity in the Bloch basis form, i.e.
\begin{equation}
	\mathbbm{1}
		= \sum_{l=v,c}\int \dd \vb{K}' \dyad{\vb{K}',l}{\vb{K}',l} ,
\end{equation}
we can rewrite Eq.~\eqref{Eq:App:Sch:proj} as
\begin{equation}\label{Eq:App:Sch:Int}
	i\hbar \pdv{}{t}\ket{\Phi_m(\vb{K},t)}
		= \sum_{l=v,c}\int \dd \vb{K'} 
			\Big[
				\mathsf{e}\mel{\vb{K},m}{\hat{\bar{r}}_{i,\text{ter}}(t)}{\vb{K}',l}\hat{E}(t) \ket{\Phi_l(\vb{K}',t)}
				-i\mathsf{e}\hbar^{-1}\mel{\vb{K},m}{\hat{\bar{v}}_{i,\text{tra}}(t)}{\vb{K}',l}\hat{A}(t) \ket{\Phi_l(\vb{K}',t)}
			\Big].
\end{equation}

Let us now proceed with the evaluation of the matrix elements within the aforementioned expression. To facilitate this, we consider that, in line with Eq.~\eqref{Eq:scl:Sch}, we can express
\begin{equation}
	U_{\text{sc}}(t,t_0)\ket{\vb{K},m}
		= b_v(\vb{K},t;m) \ket{\vb{K},v} +  b_c(\vb{K},t;m) \ket{\vb{K},c},
\end{equation}
such that for the interband contributions we get
\begin{equation}
	\begin{aligned}\label{Eq:App:Eq:inter:curr}
			&\int \dd \vb{K}' \mel{\vb{K},m}{\hat{\bar{r}}_{i,\text{ter}}(t)}{\vb{K}',l} \ket{\Phi_l(\vb{K}',t)}
			\\ &\hspace{2cm}
			= \int \dd \vb{K}'
				\big[
					b^*_v(\vb{K},t;m) \bra{\vb{K},v} +  b^*_c(\vb{K},t;m) \bra{\vb{K},c}
				\big]
				\hat{r}_{i,\text{ter}}
				\big[
					b_v(\vb{K},t;l) \ket{\vb{K},v} +  b_c(\vb{K},t;l) \ket{\vb{K},c}
				\big]\ket{\Phi_l(\vb{K}',t)}
			\\ &\hspace{2cm}
			= \sum_{\mathsf{i}=v,c}
				\sum_{\mathsf{j}=v,c}
					b^*_{\mathsf{i}}(\vb{K},t;m)
						b_{\mathsf{j}}(\vb{K},t;l)
							d_{\mathsf{ij}}
								\big(
									\vb{K} + \mathsf{e}\boldsymbol{\mathcal{E}}_iA_{\text{cl}}(t)
								\big)
								\ket{\Phi_l(\vb{K},t)}
			\\ &\hspace{2cm}
			\equiv
				\mathsf{e}^{-1}M^{(\text{ter})}_{m,l}(\vb{K},t)\ket{\Phi_l(\vb{K},t)},
	\end{aligned}
\end{equation}
where, in going from the first to the second equality, we have used Eq.~\eqref{Eq:inter:els} along with the fact that $\ket{\vb{K},m} = e^{-i\mathsf{e}A_{\text{cl}}(t)\hat{r}_{i,\text{tra}}}\ket{\phi_{\vb{k},m}}$, with $\vb{k} = \vb{K} + \boldsymbol{\mathcal{E}}_i A_{\text{cl}}(t)$. it is worth noting that the matrix element $M^{(\text{ter})}_{m,l}$ also encompasses the Berry connection terms; however, in our numerical computations, these terms are omitted due to their absence in the case of ZnO when excited along the $\Gamma-M$ crystal direction (see e.g. Ref~\cite{jiang_role_2018}).

On the other hand, to calculate the intraband matrix elements, we break it down into two parts. Since $\hat{v}_{i,\text{ter}}$ is defined as a commutator, we examine each element within the commutator individually. For one of these, we obtain
\begin{equation}\label{Eq:App:Mel:tra:I}
	\begin{aligned}
	&\int \dd \vb{K}' \mel{\vb{K},m}{\hat{\bar{r}}_{i,\text{tra}}(t)\hat{\bar{H}}_{\text{cr}}(t)}{\vb{K}',l} \ket{\Phi_l(\vb{K}',t)}
	\\ &\hspace{1cm}
	= \int \dd \vb{K}'
		\big[
			b^*_v(\vb{K},t;m) \bra{\vb{K},v} +  b^*_c(\vb{K},t;m) \bra{\vb{K},c}
		\big]
		\hat{r}_{i,\text{tra}}\hat{H}_{\text{cr}}(t)
		\big[
			b_v(\vb{K},t;l) \ket{\vb{K},v} +  b_c(\vb{K},t;l) \ket{\vb{K},c}
		\big]\ket{\Phi_l(\vb{K}',t)}
	\\ &\hspace{1cm}
	= \int \dd \vb{K}'
		\big[
			b^*_v(\vb{K},t;m) \bra{\vb{K},v} +  b^*_c(\vb{K},t;m) \bra{\vb{K},c}
		\big]
		\hat{r}_{i,\text{tra}}
		\\&\hspace{3cm}
		\times
		\big[
		E_v\big(\vb{K}'+\mathsf{e}\boldsymbol{\mathcal{E}_i}A_{\text{cl}}(t)\big)
			b_v(\vb{K},t;l) \ket{\vb{K},v} +
		E_c\big(\vb{K}'+\mathsf{e}\boldsymbol{\mathcal{E}_i}A_{\text{cl}}(t)\big) b_c(\vb{K},t;l) \ket{\vb{K},c}
		\big]\ket{\Phi_l(\vb{K}',t)}
	\\&\hspace{1cm}
	= \sum_{\mathsf{i}=c,v}
		\int \dd \vb{K}'
					b^*_{\mathsf{i}}(\vb{K},t;m)
						\mel{\vb{K},m}{\hat{r}_{i,\text{tra}}}{\vb{K}',l}
							b_{\mathsf{i}}(\vb{K}',t;l)E_{\mathsf{i}}\big(\vb{K}'+\mathsf{e}\boldsymbol{\mathcal{E}_i}A_{\text{cl}}(t)\big)\ket{\Phi_l(\vb{K}',t)}
	\\&\hspace{1cm}
	= i\hbar\sum_{\mathsf{i}=c,v}
		b^*_{\mathsf{i}}(\vb{K},t;m)
			\pdv{}{K_i}
				\big[
					b_{\mathsf{i}}(\vb{K},t;l)E_{\mathsf{i}}\big(\vb{K}+\mathsf{e}\boldsymbol{\mathcal{E}_i}A_{\text{cl}}(t)\big)\ket{\Phi_l(\vb{K},t)}
				\big],
	\end{aligned}
\end{equation}
where in this expression: (1) to transition from the second to the third equality, we consider the diagonal nature of the $\hat{r}_{i,\text{tra}}$ elements with respect to the bands, and (2) to move from the third to the fourth equality, we utilize the relationship defined in Eq.~\eqref{Eq:intra:els}. Applying a similar analysis, we get the following for the remaining term in the commutator
\begin{equation}\label{Eq:App:Mel:tra:II}
	\begin{aligned}
		&\int \dd \vb{K}' \mel{\vb{K},m}{\hat{\bar{H}}_{\text{cr}}(t)\hat{\bar{r}}_{i,\text{tra}}(t)}{\vb{K}',l} \ket{\Phi_l(\vb{K}',t)}
			= i\hbar\sum_{\mathsf{i}=c,v}
				b^*_{\mathsf{i}}(\vb{K},t;m)
				E_{\mathsf{i}}\big(\vb{K}+\mathsf{e}\boldsymbol{\mathcal{E}_i}A_{\text{cl}}(t)
				\pdv{}{K_i}
				\big[
					b_{\mathsf{i}}(\vb{K},t;l)\big)\ket{\Phi_l(\vb{K},t)}
				\big],
	\end{aligned}
\end{equation}
and combining Eqs.~\eqref{Eq:App:Mel:tra:I} and \eqref{Eq:App:Mel:tra:II}, we end up with
\begin{equation}\label{Eq:App:Eq:intra:curr}
	\begin{aligned}
		\int \dd \vb{K}' \mel{\vb{K},m}{\hat{\bar{v}}_{i,\text{tra}}(t)}{\vb{K}',l}
			&=i\hbar\sum_{\mathsf{i}=c,v}
				b^*_{\mathsf{i}}(\vb{K},t;m)
					b_{\mathsf{i}}(\vb{K},t;l)
						\pdv{}{K_i}
						\big[
							E_{\mathsf{i}}\big(\vb{K}+\mathsf{e}\boldsymbol{\mathcal{E}_i}A_{\text{cl}}(t)\big)
						\big]
					\ket{\Phi_l(\vb{K},t)}
			\\&
			\equiv i\hbar \mathsf{e}^{-1}M_{m,l}^{(\text{tra})}(\vb{K},t)\ket{\Phi_l(\vb{K},t)}.
	\end{aligned}
\end{equation}

The expressions we have found in Eqs.~\eqref{Eq:App:Eq:inter:curr} and \eqref{Eq:App:Eq:intra:curr} are directly linked to the inter- and intraband currents given in Eqs.~\eqref{Eq:inter:curr} and \eqref{Eq:intra:curr}. Specifically, the former corresponds to the time-dependent interband polarization while the second with the intraband current. However, the main difference between them is that the first ones have been computed under a noiseless scenario, while the second using the SBE formalism and introducing dephasing effects. In our numerical analysis, we take into account dephasing effects phenomenologically on the quantum optical state when computing $M_{m,l}^{(\text{ter})}$ and $M_{m,l}^{(\text{tra})}$, which we do by using the SBE approach. More specifically, we can express the $M^{(\text{ter})}_{v,v}$ and $M^{(\text{tra})}_{v,v}$ elements, which will play a fundamental influence in the equations below as
\begin{align}
    &M^{\text{ter}}_{v,v}
        = \pi(\vb{K},t)d_{cv}\big(\vb{K} + \mathsf{e}\boldsymbol{\mathcal{E}_i} A_{\text{cl}(t)}(t)\big) + \text{c.c.}
    \\&M^{\text{tra}}_{v,v}
        = \sum_{m = c,v} n_m(\vb{K},t)\pdv{}{K_i}
						\big[
							E_{\mathsf{i}}\big(\vb{K}+\mathsf{e}\boldsymbol{\mathcal{E}_i}A_{\text{cl}}(t)\big)
						\big],
\end{align}
where the $n_m(\vb{K},t)$ and $\pi(\vb{K},t)$ are obtained by numerically solving Eqs.~\eqref{Eq:semic:pop} and \eqref{Eq:semic:coh} in the main text, with the dephasing time $T_2$ used as a parameter. It is a reasonable assumption to make a weak-coupling limit approximation between the quantum optical degrees of freedom and the time-dependent interband and intraband operators since $g(\omega_L) \ll 1$. We note that under this approximation, the results obtained in this work and the ones obtained by solving the
corresponding Liouville equation should converge~\cite{wang_quantum_2021}, as in both cases we get that the final joint state of the electron and the electromagnetic field are decoupled. 

By introducing these expressions into Eq.~\eqref{Eq:App:Sch:Int} and explicitly distinguishing between projections onto the valence and conduction bands, we obtain
\begin{align}
	& i\hbar \pdv{}{t}\ket{\Phi_v(\vb{K},t)}
			= \big[
					M^{(\text{ter})}_{v,v}(\vb{K},t)\hat{E}(t)
					 +M^{(\text{tra})}_{v,v}\hat{A}(t)
				\big]\ket{\Phi_v(\vb{K},t)}
				+
				 \big[
				M^{(\text{ter})}_{v,c}(\vb{K},t)\hat{E}(t)
				+M^{(\text{tra})}_{v,c}\hat{A}(t)
				\big]\ket{\Phi_c(\vb{K},t)},
				\label{Eq:App:QO:v}
	\\
	& i\hbar \pdv{}{t}\ket{\Phi_c(\vb{K},t)}
			= \big[
					M^{(\text{ter})}_{c,v}(\vb{K},t)\hat{E}(t)
					+M^{(\text{tra})}_{c,v}\hat{A}(t)
				\big]\ket{\Phi_v(\vb{K},t)}
				+
				\big[
						M^{(\text{ter})}_{c,c}(\vb{K},t)\hat{E}(t)
						+M^{(\text{tra})}_{c,c}\hat{A}(t)
				\big]\ket{\Phi_c(\vb{K},t)}.
				\label{Eq:App:QO:c}
\end{align}

\begin{figure*}
	\centering
	\includegraphics[width = 0.75\textwidth]{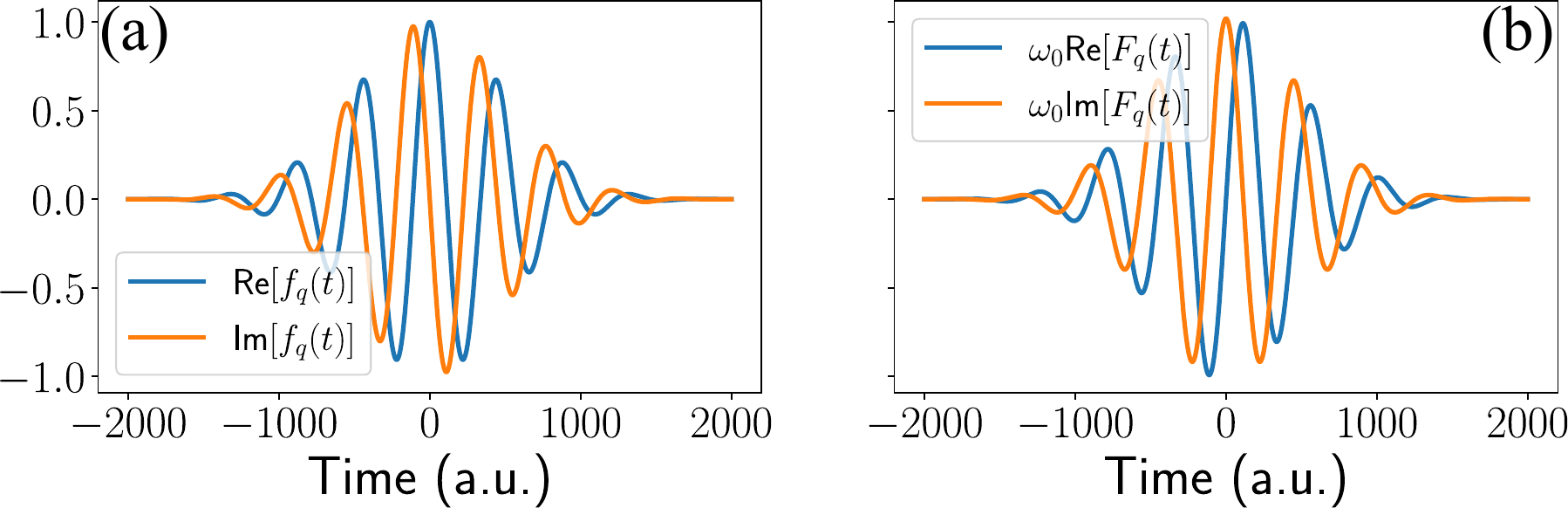}
	\caption{Real (in blue) and imaginary (in orange) parts of the envelope functions appearing in the definition of the (a) electric field operator and (b) vector potential operator when $\omega_q = \omega_L$ (similar behaviors are obtained for higher frequencies). In the limits $t \to \pm \infty$, both functions tend to zero.}
	\label{Fig:App:Envelopes}
\end{figure*}

Each of these differential equations have both a homogeneous and an inhomogeneous component. As a result, their solutions can be expressed as the sum of the solution to the corresponding homogeneous equation and a particular solution to the inhomogeneous one. Consequently, we will initially examine the homogeneous component, which is generally described by
\begin{equation}\label{Eq:App:Diff:Hom}
	i \hbar \pdv{}{t} \ket{\Phi_{\mathsf{i},\text{hom}}(\vb{K},t)}
		=  \big[
					M^{(\text{ter})}_{\mathsf{i},\mathsf{i}}(\vb{K},t)\hat{E}(t)
				+M^{(\text{tra})}_{\mathsf{i},\mathsf{i}}\hat{A}(t)
		\big]\ket{\Phi_{\mathsf{i},\text{hom}}(\vb{K},t)},
\end{equation}
and taking into account the definition of the electric field and vector potential operators, i.e.
\begin{align}
	&\hat{E}(t) 
		= -if(t) \sum^{q_c}_{q=1} \text{g}(\omega_q) 
			\big( 
				\hat{a}_q^\dagger e^{i\omega_q t}
				- \hat{a}_q e^{-i\omega_q t}
			\big),\label{Eq:App:E:def}
	\\
	& \hat{A}(t) = - \int \dd t \hat{E}(t)
						= i
							 \sum^{q_c}_{q=1} \text{g}(\omega_q) 
							\big[ 
								F_q(t)\hat{a}_q^\dagger
								-F^*_q(t) \hat{a}_q
							\big],\label{Eq:App:A:def}
\end{align}
where in Eq.~\eqref{Eq:App:A:def} we have defined $F(t) \equiv \int \dd t f_q(t)$ with $f_q(t) \equiv f(t)e^{i\omega_q t}$. In Fig.~\ref{Fig:App:Envelopes} we show how these two functions behave when setting $\omega_q = \omega_L$. Thus, by introducing these expressions in Eq.~\eqref{Eq:App:Diff:Hom}, we get
\begin{equation}
	\begin{aligned}
		i\hbar \pdv{}{t} \ket{\Phi_{\mathsf{i},\text{hom}}(\vb{K},t)}
	&=  i\sum_{q=1}^{q_c}g(\omega_q)
			\bigg[
				\Big(
					-M_{\mathsf{i},\mathsf{i}}^{(\text{ter})}(\vb{K},t)f_q(t)
					+M_{\mathsf{i},\mathsf{i}}^{(\text{tra})}(\vb{K},t)F_q(t)
				\Big)\hat{a}^\dagger_q
		\\&\hspace{1.8cm}
				-
				\Big(
					-M_{\mathsf{i},\mathsf{i}}^{(\text{ter})}(\vb{K},t)f^*_q(t)
					+M_{\mathsf{i},\mathsf{i}}^{(\text{tra})}(\vb{K},t)F^*_q(t)
				\Big)\hat{a}_q		
		\bigg]\ket{\Phi_{\mathsf{i},\text{hom}}(\vb{K},t)}.
	\end{aligned}
\end{equation}

Hence, the homogeneous part of the equation is expressed as a linear combination of photon annihilation and creation operators, which operate on distinct electromagnetic field modes. In fact, this equation bears resemblance to those derived in the analysis of HHG processes in atomic systems~\cite{lewenstein_generation_2021,rivera-dean_strong_2022,stammer_quantum_2023}, and to some extent, in molecular systems~\cite{rivera-dean_quantum_2023}. By solving the equation using a similar approach as applied in those instances, we can formulate its solution as follows:
\begin{equation}
	\ket{\Phi_{\mathsf{i},\text{hom}}(\vb{K},t)}
		= \hat{\mathcal{D}}
			\big(
				\boldsymbol{\chi}_{\mathsf{i}}(\vb{K},t,t_0)
			\big) 
			\ket{\Phi_{\mathsf{i},\text{hom}}(\vb{K},t_0)},
\end{equation}
where $\hat{\mathcal{D}}(\boldsymbol{\chi}_{\mathsf{i}}(\vb{K},t,t_0)) = \prod_{q=1}^{q_c}[e^{i\varphi_q(\vb{K},t,t_0)}\hat{D}_{q}(\chi^{(q)}_{\mathsf{i}}(\vb{K},t,t_0))]$, with $\varphi_q(\vb{K},t,t_0)$ a phase factor arising from the use of the BCH formula~\cite{Gerry__Book_2005_ch2,rivera-dean_strong_2022,stammer_quantum_2023} in order to write the solution above, and where $\chi_q(\vb{K},t,t_0)$ is given by
\begin{equation}
	\chi^{(q)}_{\mathsf{i}}\big(\vb{K},t,t_0\big)
		= \dfrac{1}{\hbar}g(\omega_q)
			\int^t_{t_0} \dd \tau
			\Big[
				-M_{\mathsf{i},\mathsf{i}}^{(\text{ter})}(\vb{K},\tau)f_q(\tau)
				+M_{\mathsf{i},\mathsf{i}}^{(\text{tra})}(\vb{K},\tau)F_q(\tau)
			\Big].
\end{equation}

If we now concentrate on Eqs.~\eqref{Eq:App:QO:v} and \eqref{Eq:App:QO:c}, and upon introducing the specific solution to the inhomogeneous equation, we can establish the subsequent recursion relation for their solutions
\begin{align}
	& \ket{\Phi_{v}(\vb{K},t)}
		= \hat{\mathcal{D}}
			\big(
				\boldsymbol{\chi}_{v}(\vb{K},t,t_0)
			\big)
			 \ket{\Phi_{v}(\vb{K},t_0)}
			 - \dfrac{i}{\hbar}
			 	\int^t_{t_0} \dd t'
			 		\hat{\mathcal{D}}
		 			\big(
			 			\boldsymbol{\chi}_{v}(\vb{K},t,t')
			 		\big)
			 		\hat{\mathcal{M}}_{v,c}(\vb{K},t')
			 		\ket{\Phi_c(\vb{K},t')},
		\\
		& \ket{\Phi_{c}(\vb{K},t)}
		= \hat{\mathcal{D}}
			\big(
				\boldsymbol{\chi}_{c}(\vb{K},t,t_0)
			\big)
			\ket{\Phi_{c}(\vb{K},t_0)}
			- \dfrac{i}{\hbar}
				\int^t_{t_0} \dd t'
					\hat{\mathcal{D}}
					\big(
						\boldsymbol{\chi}_{c}(\vb{K},t,t')
					\big)
					\hat{\mathcal{M}}_{c,v}(\vb{K},t')
					\ket{\Phi_v(\vb{K},t')},
\end{align}
where in this expression we have defined, for the sake of clarity
\begin{equation}
	\hat{\mathcal{M}}_{\mathsf{i},\mathsf{j}}(\vb{K},t')
		\equiv		
			M^{(\text{ter})}_{\mathsf{i},\mathsf{j}}(\vb{K},t)\hat{E}(t)
			+M^{(\text{tra})}_{\mathsf{i},\mathsf{j}}\hat{A}(t).
\end{equation}

By combining these two equations and considering terms up to first order in $g(\omega_q)$, we obtain the following expression after introducing the initial conditions
\begin{align}
	&\ket{\Phi_{v}(\vb{K},t)}
		= \delta(\vb{K}-\vb{K}_0)
			\hat{\mathcal{D}}
				\big(
					\boldsymbol{\chi}_{v}(\vb{K},t,t_0)
				\big)
			\bigotimes^{q_c}_{q=1}\ket{0_q}
		\\
	& \ket{\Phi_{c}(\vb{K},t)}
		= 
		- \delta(\vb{K}-\vb{K}_0)
		\dfrac{i}{\hbar}
			\int^t_{t_0} \dd t'
				\hat{\mathcal{D}}
				\big(
					\boldsymbol{\chi}_{c}(\vb{K},t,t')
				\big)
				\hat{\mathcal{M}}_{c,v}(\vb{K},t')
					\bigotimes^{q_c}_{q=1}\ket{0_q},
\end{align}
and for the joint state of the system we get (up to a normalization factor)
\begin{equation}
	\begin{aligned}
	\tPsi
		&= \sum_{m=v,c}
			\int \dd \vb{K}
				\ket{\vb{K},m}
					\ket{\Phi_{m}(\vb{K},t)}
		\\
		&= \hat{\mathcal{D}}
			\big(
				\boldsymbol{\chi}_{v}(\vb{K}_0,t,t_0)
			\big)
			\ket{\vb{K_0},v}\bigotimes^{q_c}_{q=1}\ket{0_q}
			- 
			\dfrac{i}{\hbar}
				\int^t_{t_0} \dd t'
					\hat{\mathcal{D}}
					\big(
						\boldsymbol{\chi}_{c}(\vb{K}_0,t,t')
					\big)
					\hat{\mathcal{M}}_{c,v}(\vb{K}_0,t')
						\ket{\vb{K}_0,c}
						\bigotimes^{q_c}_{q=1}\ket{0_q}.
		\end{aligned}
\end{equation}

Finally, we reverse the transformations that were previously implemented in order to get the set of differential equations we have just discussed. To be more precise, we encounter the following
\begin{equation}\label{Eq:App:Back:to:Original}
	\ket{\bar{\Psi}(t)}
		=  e^{i\mathsf{e}(A_{\text{cl}}(t) + \hat{A}(t))\hat{r}_{i,\text{tra}}/\hbar}
			\hat{U}_{\text{sc}}(t,t_0)
				\toPsi,
\end{equation}
which in the asymptotic limit $t \to \infty$, leads to
\begin{equation}\label{Eq:App:Back:to:Original:II}
		\ket{\bar{\Psi}(t)}
			=\hat{U}_{\text{sc}}(t,t_0)
				\toPsi,
\end{equation}
since, as observed in Fig.~\ref{Fig:App:Envelopes}, the contribution of the first unitary transformation appearing in Eq.~\eqref{Eq:App:Back:to:Original} tends towards the unity. This is because the envelope functions present in the the vector potential's definition approach zero in the considered asymptotic limit, as depicted in Fig.~\ref{Fig:App:Envelopes}~(b). Shifting our focus to processes where the electron ultimately resides in the valence band --thus applying the projector $\hat{P}_v = \int \dd \vb{K} \dyad{\vb{K},v}$ to Eq.~\eqref{Eq:App:Back:to:Original:II}-- and tracing over the electronic degrees of freedom under while assuming that the probability of finding in the valence band electrons initially born in the conduction band is small, we arrive at
\begin{equation}
	\ket{\Phi_v(\vb{K},t)}
		\approx 
			\hat{\mathcal{D}}
			\big(
				\boldsymbol{\chi}_{v}(\vb{K}_0,t,t_0)
			\big)
			\bigotimes^{q_c}_{q=1}\ket{0_q},
\end{equation}
similar to what happens in atomic systems~\cite{lewenstein_generation_2021,rivera-dean_strong_2022,stammer_quantum_2023}. It is worth noting that incorporating terms like $\langle \vb{K},v\vert U_{\text{sc}}(t,t_0)\vert \vb{K}',c\rangle$, would introduce entanglement between the electronic and electromagnetic field states. This was also recently studied in Ref.~\cite{rivera-dean_entanglement_2023}, although under a more convenient Wannier perspective where one could treat the electronic states in a discrete manner, instead of the less convenient continuous representation described in this text. However, it was shown there that for ZnO materials the amount of light-matter entanglement was almost negligible.

\end{document}